\documentclass[a4paper]{iopart}

\usepackage[dvips]{graphicx}
\usepackage{times}
\usepackage{latexsym}
\usepackage{amssymb}
\usepackage{color}

\newcommand{\tg}{\tilde\gamma}
\newcommand{\tG}{\tilde\Gamma}
\newcommand{\tA}{\tilde A}

\newcommand{\ie}{\textit{i.e.}~}

\begin{document}

\title[Improved formulation of the relativistic hydrodynamics eqs. in
  2D Cartesian coords.]{An improved formulation of the relativistic
  hydrodynamics equations in 2D Cartesian coordinates}

\author{Thorsten Kellerman$^{1}$, Luca Baiotti$^{1,2}$, Bruno
  Giacomazzo$^{1}$ and Luciano Rezzolla$^{1,3,4}$}

\address{$^1$ Max-Planck-Institut f\"ur Gravitationsphysik, Albert Einstein 
Institut, 
Golm, Germany}

\address{$^2$ Graduate School of Arts and Sciences, University of Tokyo,
Komaba, Meguro-ku, Tokyo, 153-8902, Japan}

\address{$^3$ Department of Physics, Louisiana State University, Baton
  Rouge, LA 70803 USA}

\address{$^4$ INFN, Sezione di Trieste, 
Trieste, Italy}

\begin{abstract}
A number of astrophysical scenarios possess and preserve an overall
cylindrical symmetry also when undergoing a catastrophic and nonlinear
evolution. Exploiting such a symmetry, these processes can be studied
through numerical-relativity simulations at smaller computational
costs and at considerably larger spatial resolutions. We here present
a new flux-conservative formulation of the relativistic hydrodynamics
equations in cylindrical coordinates. By rearranging those terms in
the equations which are the sources of the largest numerical errors,
the new formulation yields a global truncation error which is one or
more orders of magnitude smaller than those of alternative and
commonly used formulations. We illustrate this through a series of
numerical tests involving the evolution of oscillating spherical and
rotating stars, as well as shock-tube tests.
\end{abstract}
\maketitle

\section{Introduction}

Numerical simulations assuming and enforcing axisymmetry are
particularly useful to study at higher resolution and smaller
computational costs those astrophysical scenarios whose evolution is
expected to possess and preserve such a symmetry. On the other side,
the numerical solution of systems of equations expressed in
coordinates adapted to the symmetry has often posed serious
difficulties, because of the coordinate singularity present on the
symmetry axis. The \textit{``cartoon''} method, proposed by Alcubierre
{\it et al.}~\cite{Alcubierre99a_nourl}, allows to exploit the
advantages of reduced computational resource requirements, while
adopting Cartesian coordinates, which are non-singular.

The \textit{``cartoon''} method proves particularly useful in the numerical
evolution of smooth functions, like the metric quantities of the
Einstein equations. However, because of the interpolations necessary
to impose the axisymmetric conditions on a Cartesian grid, the
\textit{``cartoon''} approach is not considered to be accurate enough to
describe the shocks which generically develop when matter is
present. As a consequence, general-relativistic codes employing the
\textit{``cartoon''} method have adopted cylindrical coordinates for the
evolution of the matter (and magnetic field)
variables~\cite{Shibata00a,Shibata:2003iy,Montero:2004,Shibata05b,Montero2008}.
All the cited works adopt the same formulation for the hydrodynamical
equations in cylindrical coordinates. In the present article, we
propose a slightly different formulation, which has proven to reduce
the numerical errors, especially in the vicinity of the symmetry axis.

More specifically, we have written the \texttt{Whisky2D} code, which
solves the general-relativistic hydrodynamics equations in a
flux-conservative form and in cylindrical coordinates. This of course
brings in $1/r$ singular terms, which must be dealt with
appropriately. In the above-referenced works, the flux operator is
expanded and the $1/r$ terms, not containing derivatives, are moved to
the right hand side of the equation (the source term), so that the
left hand side assumes a form identical to the one of the
three-dimensional (3D) Cartesian formulation. We call this the {\it
  standard formulation}. An other possibility is not to split the flux
operator and to redefine the conserved variables, via a multiplication
by $r$. We call this the {\it new formulation}.  The new equations are
solved with the same methods as in the Cartesian case. From a
mathematical point of view, one would not expect differences between
the two ways of writing the differential operator, but, of course, a
difference is present at the numerical level. Our tests show that the
new formulation yields results with a global truncation error which is
one or more orders of magnitude smaller than those of alternative and
commonly used formulations.

Here we perform a series of tests to ascertain the convergence
behaviour of the two formulations. We then show that the new
formulation produces results which are generally more accurate, with a
truncation error which can be several orders of magnitude smaller.

The paper in organized as follows. In Section~\ref{EinsteinEq} we
remind the essentials of the \textit{``cartoon''} approach for the evolution of
the geometrical variables, while in Section~\ref{hydro_eqs} we review
the flux-conservative formulation of relativistic hydrodynamics. We
write down the relativistic flux-conservative hydrodynamics equations
for axisymmetric formulations and we illustrate the two possible ways
to write the singular term. In Section~\ref{numtests}, we present
several tests that compare the two formulations. We begin with the
conservation of rest mass and angular momentum in the Cowling
approximation and in full-spacetime evolution. Then the
eigenfrequencies of uniformly rotating neutron-star models are compared
with the results of a perturbative code. The last test examines the
differences between the two formulations with respect to an analytic
solution of an extreme shock case, which mimics the reflection of a
cold and very fast gas at the symmetry axis.

We have used a spacelike signature $(-,+,+,+)$, with Greek indices
running from 0 to 3, Latin indices from 1 to 3 and the standard
convention for the summation over repeated indices. Unless explicitly
stated, all the quantities are expressed in the system of
dimensionless units in which $c=G=M_\odot=1$.

\section{Evolution of the Einstein equations}
\label{EinsteinEq}

The logical and algorithmic structures of the \texttt{Whisky2D} code
presented here follow closely the ones of the
\texttt{CCATIE}~\cite{Pollney:2007ss} and \texttt{Whisky}
codes~\cite{Baiotti03a}, which solve the same set of equations in 3D
and using Cartesian coordinates. In what follows we provide only a
brief overview of the set of equations for the evolution of the fields
(within the \textit{``cartoon''}
prescription~\cite{Alcubierre99a_nourl}) and for the evolution of the
fluid variables, referring the interested reader to
refs.~\cite{Pollney:2007ss,Baiotti04,Baiotti08} for a more detailed
discussion. As for the other codes mentioned above, also
\texttt{Whisky2D} is based on the Cactus Computational
Toolkit~\cite{Goodale02a}.

More specifically, we evolve a conformal-traceless ``$3+1$''
formulation of the Einstein
equations~\cite{Nakamura87,Shibata95,Baumgarte99}, in which the
spacetime is decomposed into 3D spacelike slices, described by a
metric $\gamma_{ij}$, its embedding in the full spacetime, specified
by the extrinsic curvature $K_{ij}$, and the gauge functions $\alpha$
(lapse) and $\beta^i$ (shift), which specify a coordinate frame
(see~\cite{York79} for a general description of the $3+1$ split). The
particular system which we evolve transforms the standard ADM
variables as follows. The 3-metric $\gamma_{ij}$ is conformally
transformed via
\begin{equation}
  \label{eq:def_g}
  \Phi = \frac{1}{12}\ln \det \gamma_{ij}, \qquad
  \tilde{\gamma}_{ij} = e^{-4\Phi} \gamma_{ij},
\end{equation}
and the conformal factor $\Phi$ evolved as an independent variable,
whereas $\tilde{\gamma}_{ij}$ is subject to the constraint
$\det \tilde{\gamma}_{ij} = 1$. The extrinsic curvature is
subjected to the same conformal transformation and its trace
$\tr K_{ij}$ is evolved as an independent variable. That is, in place of
$K_{ij}$ we evolve:
\begin{equation}
  \label{eq:def_K}
  K \equiv \tr K_{ij} = g^{ij} K_{ij}, \qquad
  \tilde{A}_{ij} = e^{-4\Phi} (K_{ij} - \frac{1}{3}\gamma_{ij} K),
\end{equation}
with $\tr\tilde{A}_{ij}=0$. Finally, new evolution variables
\begin{equation}
  \label{eq:def_Gamma}
  \tilde{\Gamma}^i = \tilde{\gamma}^{jk}\tilde{\Gamma}^i_{jk}
\end{equation}
are introduced, defined in terms of the Christoffel symbols of
the conformal 3-metric.

The Einstein equations specify a well known set of evolution equations
for the listed variables and are given by 
\begin{eqnarray}
  \label{eq:evolution}
\fl
  && \hskip 1.0cm 
 (\partial_t - {\cal L}_\beta)\; \tg_{ij}  = -2 \alpha \tA_{ij},  \\
\fl
  && \hskip 1.0cm 
 (\partial_t - {\cal L}_\beta)\; \Phi  = - \frac{1}{6} \alpha K, \\
\fl
  &&\hskip 1.0cm 
 (\partial_t - {\cal L}_\beta)\; \tA_{ij}  = e^{-4\Phi} [ - D_i D_j \alpha 
   + \alpha (R_{ij} - 8 \pi S_{ij})]^{TF} 
+ \alpha (K \tA_{ij} - 2 \tA_{ik} \tA^k{}_j), \\
\fl
  && \hskip 1.0cm 
 (\partial_t - {\cal L}_\beta)\; K  = - D^i D_i \alpha
   + \alpha\left[\tA_{ij} \tA^{ij} + \frac{1}{3} K^2 + 
   4\pi (\rho_{_{\rm  ADM}}+S)\right], \\
\fl
  && \hskip 1.0cm 
 \partial_t \tG^i  = \tilde\gamma^{jk} \partial_j\partial_k \beta^i
    + \frac{1}{3} \tilde\gamma^{ij}  \partial_j\partial_k\beta^k
    + \beta^j\partial_j \tilde\Gamma^i
   - \tilde\Gamma^j \partial_j \beta^i 
   + \frac{2}{3} \tilde\Gamma^i \partial_j\beta^j \nonumber \\
\fl
   && \hskip 2.0cm 
   - 2 \tilde{A}^{ij} \partial_j\alpha
   + 2 \alpha \left( 
   \tilde{\Gamma}^i{}_{jk} \tilde{A}^{jk} + 6 \tilde{A}^{ij}
   \partial_j \Phi 
- \frac{2}{3} \tg^{ij} \partial_j K - 8 \pi \tg^{ij} S_j\right),
\end{eqnarray}
where $R_{ij}$ is the three-dimensional Ricci tensor, $D_i$ is the
covariant derivative associated with the three metric $\gamma_{ij}$,
``TF'' indicates the trace-free part of tensor objects and $
{\rho}_{_{\rm ADM}}$, $S_j$ and $S_{ij}$ are the matter source terms
defined as
\begin{eqnarray}
\rho_{_{\rm ADM}}&\equiv n_\alpha n_\beta T^{\alpha\beta}, \nonumber \\ 
S_i&\equiv -\gamma_{i\alpha}n_{\beta}T^{\alpha\beta}, \\
S_{ij}&\equiv \gamma_{i\alpha}\gamma_{j\beta}T^{\alpha\beta}, \nonumber
\end{eqnarray}
where $n_\alpha\equiv (-\alpha,0,0,0)$ and $T^{\alpha\beta}$ is the
stress-energy tensor for a perfect fluid (see Section~\ref{hydro_eqs}).

Four elliptic constraint equations, which are usually
referred to as Hamiltonian and momentum constraints,
\begin{eqnarray}
\fl
  \label{eq:einstein_ham_constraint}
&&    \hskip 1.0cm 
\mathcal{H} \equiv R^{(3)} + K^2 - K_{ij} K^{ij} 
  - 16\pi\rho_{_{\rm ADM}}= 0\,, \\
\fl
  \label{eq:einstein_mom_constraints}
&&   \hskip 1.0cm 
 \mathcal{M}^i \equiv D_j(K^{ij} - \gamma^{ij}K)  - 8\pi S^i = 0\,,
\end{eqnarray}
should be satisfied within each spacelike slice. Here $R^{(3)}=R_{ij}
\gamma^{ij}$ is the Ricci scalar on a 3D timeslice.  Additional
constraints are given by
\begin{eqnarray}
  \fl
  \hskip 1.0cm 
  \det \tilde{\gamma}_{ij}  = 1, 
    \hskip 2.0cm 
  \tr \tilde{A}_{ij}  = 0,
    \hskip 2.0cm 
  \tilde{\Gamma}^i  = \tilde{\gamma}^{jk}\tilde{\Gamma}^i_{jk}\, ,
   \label{eq:Gamma_def}
\end{eqnarray}
with the last two equations of (\ref{eq:Gamma_def}) being enforced
algebraically.  The remaining constraint in~(\ref{eq:Gamma_def}) and the constraints $\mathcal{H}$
and $\mathcal{M}^i$ are not actively enforced
and can be used as monitors of the accuracy of our numerical solution.

We specify the gauges in terms of the standard ADM lapse function,
$\alpha$, and shift vector, $\beta^a$~\cite{misner73}.  We evolve the
lapse according to the ``$1+\log$'' slicing condition~\cite{Bona94b}:
\begin{equation}
  \partial_t \alpha - \beta^i\partial_i\alpha 
    = -2 \alpha (K - K_0),
  \label{eq:one_plus_log}
\end{equation}
where $K_0$ is the initial value of the trace of the extrinsic
curvature and equals zero for the maximally sliced initial data we
consider here.  The code uses a hyperbolic $\tilde{\Gamma}$-driver
condition~\cite{Alcubierre02a}
\begin{eqnarray}
\label{gamma-driver}
  \partial_t \beta^i - \beta^j \partial_j  \beta^i & = & \frac{3}{4} \alpha B^i\,,
  \\
  \partial_t B^i - \beta^j \partial_j B^i & = & \partial_t \tilde\Gamma^i 
    - \beta^j \partial_j \tilde\Gamma^i - \eta B^i\,,
\end{eqnarray}
where $\eta$ is a parameter which acts as a damping coefficient (see
discussion in ref.~\cite{Loeffler06a}).

Two routes are possible when solving numerically the Einstein
equations in axisymmetric spacetimes. One route consists in using
coordinates that exploit the symmetry and enforce its preservation
already at a mathematical level, such as cylindrical coordinates. This
advantage is counterbalanced by the fact that such coordinates are
usually singular somewhere (\textit{e.g.,} on the axis for cylindrical
coordinates) and that regularization conditions are therefore
necessary (see~\cite{Rinne-Stewart-2005,Ruiz:2008} and references
therein for a recent discussion).

The second route consists, instead, in using Cartesian coordinates and in exploiting
the fact that these coincide with the cylindrical ones in one plane,
namely the $(x,z)$ plane (for concreteness we will assume hereafter
that the Cartesian and the cylindrical $z$-axes coincide). The chief
advantages of this approach, which is usually referred to as the
\textit{``cartoon''} method~\cite{Alcubierre99a_nourl}, are the absence of the need of regularization
conditions and the easiness of implementation, through a simple
dimensional reduction from fully 3D codes in Cartesian
coordinates. However, these advantages are counterbalanced by
at least two disadvantages. The first one is that the method still
essentially requires the use of a 3D domain covered with Cartesian
coordinates, although one of the three dimensions, namely the
$y$-direction, has a very small extent. The second one is that, in order
to compute the spatial derivatives in the $y$-direction appearing in
the Einstein equations, a number of high-order interpolations onto the
$x$-axis are necessary (see discussion below) and these can amount to
a significant portion of the time spent for each evolution to the new
timelevel. In practice, the spatial derivatives in the $y$-direction
are computed exploiting the fact that all quantities are
constant on cylinders and thus the value of a variable $\Psi$ at a
generic position $({x},{y},{z})$ off the $(x,z)$ plane can be computed
from the corresponding value $\Psi(\tilde{x},0,\tilde{z})$ on the
$(x,z)$ plane, where
\begin{equation}
\tilde{x} = ({x} + {y})^{1/2}\,, \qquad 
\tilde{z} = {z}\,.
\end{equation}
Clearly, since the solution of the evolution equations is computed
only on the $(x,z)$ plane, interpolations (with truncation errors
smaller than that of the finite-difference operators) are needed at
all the positions $(\tilde{x},y=0,\tilde{z})$.

Overall the \textit{``cartoon''} method represents the choice for many codes
and it has been implemented with success in many applications,
\textit{e.g.,}~\cite{Alcubierre99a_nourl,Shibata00a,Shibata:2003iy,Montero:2004,Pretorius:2004jg,Duez05MHD0,Shibata05b,Montero2008}
to cite a few.

\bigskip

\section{Evolution of the relativistic hydrodynamics equations}
\label{hydro_eqs}

An important feature of multidimensional non-vacuum numerical-relativity codes that solve the
coupled Einstein--hydrodynamics equations in Cartesian coordinates is the adoption of a
\textit{conservative} formulation of the hydrodynamics equations~\cite{Marti91,Banyuls97}. In such a
formulation, the set of conservation equations for the stress-energy tensor $T^{\mu\nu}$ and for the
matter current density $J^\mu$, that is
\begin{eqnarray}
\label{continuity}
&&\nabla_\mu J^\mu = 0\,, \\
\label{energymom_cons}
&&\nabla_\mu T^{\mu\nu} = 0\,,
\end{eqnarray}
is written in a hyperbolic, first-order ``flux-conservative'' form
of the type~\cite{Leveque92}
\begin{equation}
\label{eq:consform1}
\partial_t {\mathbf q} + 
        \partial_i {\mathbf f}^{(i)} ({\mathbf q}) = 
        {\mathbf s} ({\mathbf q})\,,
\end{equation}
where ${\mathbf f}^{(i)} ({\mathbf q})$ and ${\mathbf s}({\mathbf q})$
are the flux vectors and source terms, respectively~\cite{Font03}.  Note
that the right-hand side (the source terms) depends only on the metric,
on its first derivatives and on the stress-energy tensor. Furthermore,
while the system (\ref{eq:consform1}) is not strictly hyperbolic,
strong hyperbolicity is recovered in a flat spacetime, where ${\mathbf s}
({\mathbf q})=0$.

As shown by~\cite{Banyuls97}, in order to write the system
(\ref{continuity})--(\ref{energymom_cons}) in the form of system
(\ref{eq:consform1}), the \textit{primitive} hydrodynamical variables
({\it i.e.} the rest-mass density $\rho$, the pressure $p$ measured in
the rest-frame of the fluid, the fluid 3-velocity $v^i$ measured by a
local zero-angular momentum observer, the specific internal energy
$\epsilon$ and the Lorentz factor $W$) are mapped to the so called
\textit{conserved} variables \mbox{${\mathbf q} \equiv (D, S^i,
  \tau)$} via the relations
\begin{eqnarray}
  \label{eq:prim2con}
   D &\equiv& \sqrt{\gamma}\rho W\,, \nonumber\\
   S^i &\equiv& \sqrt{\gamma} \rho h W^2 v^i\,,  \\
   \tau &\equiv& \sqrt{\gamma}\left( \rho h W^2 - p\right) - D\,, \nonumber
\end{eqnarray}
where $h \equiv 1 + \epsilon + p/\rho$ is the specific enthalpy and
\hbox{$W \equiv (1-\gamma_{ij}v^i v^j)^{-1/2}$}. 

The advantage of a flux-conservative formulation is that it allows to
use high-resolution shock-capturing (HRSC) schemes, which are based on
Riemann solvers and which are essential for a correct representation
of shocks. This is particularly important in astrophysical
simulations, where large shocks are expected.  In this approach, all
variables ${\bf q}$ are represented on the numerical grid by
cell-integral averages. The function is then {\it reconstructed}
within each cell, usually through piecewise polynomials, in a way that
preserves the conservation of the variables ${\bf q}$. This gives two
values at each cell boundary, which are then used as initial data for
the (approximate) Riemann problem, whose solution gives the flux
through the cell boundary.

As in the \texttt{Whisky} code, the evolution equations are here
integrated in time using the method of line~\cite{Toro99}, which
reduces the partial differential equations~(\ref{eq:consform1}) to a
set of ordinary differential equations that can be evolved using
standard numerical methods, such as Runge-Kutta or the iterative
Cranck-Nicolson schemes~\cite{Teukolsky00,
  Leiler_Rezzolla06}. Furthermore, the {\tt Whisky2D} code implements
several reconstruction methods, such as Total-Variation-Diminishing
(TVD) methods, Essentially-Non-Oscillatory (ENO)
methods~\cite{Harten87} and the Piecewise-Parabolic-Method
(PPM)~\cite{Colella84}. Also, a variety of approximate Riemann solvers
can be used, starting from the Harten-Lax-van Leer-Einfeldt (HLLE)
solver~\cite{Harten83}, over to the Roe solver~\cite{Roe81} and the
Marquina flux formula~\cite{Aloy99b} (see~\cite{Baiotti03a,Baiotti04}
for a more detailed discussion).

The ability of properly evolving large gradients moving at
relativistic speeds represents one of the main motivations that make
this formulation the choice for all of the present 3D
numerical-relativity codes solving the relativistic hydrodynamics
equations on Eulerian grids (see refs.~\cite{Shibata06a, Anderson2007,
Liu01} for some of the most recent examples and
ref.~\cite{Oechslin06} for an alternative Lagrangian method). However,
when considered within the axisymmetric approach used here, the use
of a flux-conservative formulation in Cartesian coordinates may suffer
from a potentially very serious disadvantage. In fact,
the interpolations required by the \textit{``cartoon''} method may
be highly inaccurate when discontinuities in the fluid variables
appear. To confront this problem, Shibata~\cite{Shibata00a} has made the useful
suggestion of writing the relativistic hydrodynamics equations in
cylindrical coordinates, while keeping the solution of the Einstein
equations in Cartesian coordinates. This approach has the obvious
advantage that it does not require interpolation and that it exploits, at the
mathematical level, the symmetries of the system, thus guaranteeing a
better conservation of mass and angular momentum.  However,
the use of cylindrical coordinates for the evolution of
the fluid variables also comes with an undesirable property: the
coordinates are degenerate at the symmetry axis and the equations are
no longer free of singularities. As we will comment in the following
Section, this drawback can be compensated through a suitable
formulation of the equations and a proper setup of the numerical grid.

\subsection{A new formulation of the hydrodynamics equations}

As mentioned in the previous Section, following
ref.~\cite{Shibata:2003iy}, we write the relativistic hydrodynamics
equations (\ref{continuity})--(\ref{energymom_cons}) in a first-order
form in space and time using cylindrical coordinates
$(r,\phi,z)$. However, as an important difference from the approach
suggested in ref.~\cite{Shibata:2003iy}, we do not introduce source
terms that contain coordinate singularities. Rather, we 
re-define the conserved quantities in such a way to remove
the singular terms, which are the largest source of truncation error, 
also when evaluated far from the axis.

We illustrate our approach by using as a representative example the
continuity equation. This is the simplest of the five hydrodynamical
equations but already contains all the basic elements necessary to illustrate
the new formulation. We start by using the definitions for the
conserved variables~(\ref{eq:prim2con}) to write
eq.~(\ref{continuity}) generically as
\begin{equation}
\label{continuity_general}
\partial_{t}(\sqrt{\gamma} \rho W) + 
	\partial_{i}\left[\sqrt{\gamma} \rho W \left(\alpha v^{i} - 
          \beta^{i}\right)\right] = 0\,,
\end{equation}
which in cylindrical coordinates takes the form
\begin{equation}
\fl
\hskip 1.0cm
\label{continuity_cyl}
\partial_{t}(\sqrt{\tilde \gamma} \rho W) + 
	\partial_{r}\left[\sqrt{\tilde \gamma} \rho W \left(\alpha v^{r} - 
          \beta^{r}\right)\right] +
	\partial_{z}\left[\sqrt{\tilde \gamma} \rho W \left(\alpha v^{z} - 
          \beta^{z}\right)\right] = 0\,,
\end{equation}
where $\sqrt{\tilde \gamma}$ is the determinant of the 3-metric in
cylindrical coordinates and where we have enforced the condition of
axisymmetry $\partial_{\phi}=0$. Because any $\phi$-constant plane in
cylindrical coordinates can be mapped into the $(x,z)$ plane in
Cartesian coordinates, we consider equation~(\ref{continuity_cyl}) as
expressed in Cartesian coordinates and restricted to the $y=0$ plane,
\textit{i.e.,}
\begin{equation}
\fl
\hskip 2.0cm
\label{continuity_new}
\partial_{t}(x D) + 
	\partial_{x}\left[x D \left(\alpha v^{x} - \beta^{x}\right)\right] +
	\partial_{z}\left[x D \left(\alpha v^{z} - \beta^{z}\right)\right] = 
        0\,,
\end{equation}
where we have exploited the fact that for any vector of components $A^i$ on
this plane $A^{r}=A^{x},\ A^{\phi}=A^{y}$ and ${\tilde \gamma} = x^2
\gamma$, with $\gamma$ being the determinant of the 3-metric in
Cartesian coordinates. Equation~(\ref{continuity_new}) represents the
prototype of the formulation proposed here, which we will refer to
hereafter as the \textit{``new''} formulation to contrast it with the
formulation adopted so far, \textit{e.g.,} in
ref.~\cite{Shibata:2003iy}, for the solution of the relativistic
hydrodynamics equations in axisymmetry and in Cartesian
coordinates. The only, but important, difference with respect to the
\textit{``standard''} formulation is that in the latter the derivative in
the $x$-direction is written out explicitly and becomes part of the
source term ${\mathbf s} ({\mathbf q})$, \textit{i.e.,}
\begin{equation}
\fl
\hskip 2.0cm
\label{continuity_old}
\partial_{t}(D) + 
	\partial_{x}\left[D \left(\alpha v^{x} - \beta^{x}\right)\right] +
	\partial_{z}\left[D \left(\alpha v^{z} - \beta^{z}\right)\right] = 
        -\frac{D \left(\alpha v^{x} - \beta^{x}\right)}{x}\,.
\end{equation}
Even though the right-hand-side of eq.~(\ref{continuity_old}) is never
evaluated at $x=0$ (because no grid points are located at $x=0$), both the numerator and the denominator of the
right-hand-side of eq.~(\ref{continuity_old}) are very small for $x
\simeq 0$, so that small round-off errors in the evaluation of the
right-hand-side can increase the overall truncation error. Stated
differently, the right-hand-side of eq.~(\ref{continuity_old}) becomes stiff for $x
\simeq 0$ and this opens the door to the problems encountered in the
numerical solution of hyperbolic equations with stiff source terms~\cite{dumbser08}.

What was done for the continuity equation (\ref{continuity_new}) can be
extended to the other hydrodynamics equations which, for the
conservation of momentum in the $x$- and $z$-directions, take the form
\vbox{
\begin{eqnarray}
\fl
&&
\hskip -2.0cm 
\frac{1}{\alpha x \sqrt{\gamma}}
	\biggl\{\partial_{t}\left(x S_{A}\right) + 
	\partial_{x}\left[ x \left(S_{A} \left(\alpha
        v^{x} - \beta^{x}\right) + 
	\alpha \sqrt{\gamma} p \delta^{x}_{A}\right)\right] +
        \nonumber \\
&&
\hskip 4.0cm 
	\partial_{z}\left[ x \left(S_{A} \left(\alpha
        v^{z} - \beta^{z}\right) + 
	\alpha \sqrt{\gamma} p \delta^{z}_{A}\right)\right]
	\biggr\} = \nonumber \\ \nonumber \\
&&
\hskip -1.25cm \left[T^{00}\left(\frac{1}{2}\beta^{l}\beta^{m} \partial_{A}
  	\gamma_{lm}- \alpha\partial_{A}\alpha\right)+
	T^{0i}\beta^{l}\partial_{A}\gamma_{il}+
        T^{0}_{\ i}\partial_{A}\beta^{i}+\frac{1}{2}T^{lm}
        \partial_{A}\gamma_{lm}\right]\,,
\label{momentum_equation_new} 
\end{eqnarray}
}
with $A=x,z$. Similarly, the evolution of the conserved angular
momentum $S_{\phi} = x S_{y}$ is expressed as
\begin{eqnarray}
\hskip -2.0cm 
\frac{1}{\alpha x \sqrt{\gamma}} 
	\biggl\{
        \partial_{t}\left(x^2 S_{y} \right) +
        \partial_{x}\left[x^2 S_{y} \left(\alpha v^{x}-
        \beta^{x}\right)\right] + 
        \partial_{z}\left[x^2 S_{y} \left(\alpha v^{z}-
        \beta^{z}\right)\right]\biggr\} = 0\,,
\label{angular_momentum_equation} 
\end{eqnarray}
while the equation of the energy conservation is given by
\begin{eqnarray}
&&
\hskip -2.0cm
\frac{1}{\alpha x \sqrt{\gamma}}  
	\biggl\{ 
        \partial_{t}\left(x \tau\right) + 
        \partial_{x}\left[x \left(\tau \left(\alpha v^{x} - 
	\beta^{x}\right) + p v^{x}\right)\right] +
        \partial_{z}\left[x \left(\tau \left(\alpha v^{z} - 
	\beta^{z}\right) + p v^{z}\right)\right] \biggr\}
        = \nonumber\\ \nonumber \\
&&
\hskip 0.0cm
	T^{00}\left(\beta^{i}\beta^{j}K_{ij}-
	\beta^{i}\partial_{i}\alpha\right) + 
	T^{0i}\left(-\partial_{i}\alpha + 2\beta^{j}K_{ij}\right)+
	T^{ij}K_{ij}\,.
\label{energy_equation_new}
\end{eqnarray}

The changes made to the formulation are rather simple but, as we will
show in Section~\ref{numtests}, these can produce significant
improvements on the overall accuracy of the simulations with a
truncation error at least one order of magnitude
smaller for all of the tests considered. Because of its simplicity, the
changes in the new formulation of the equations can be implemented
straightforwardly in codes written using the standard formulation.

Finally, we note that both eq.~(\ref{continuity_old}) and
eq.~(\ref{continuity_new}) are written in a flux-conservative form in
the sense that the source term does not contain first-order spatial
derivatives of the conserved variables. More precisely,
eq.~(\ref{continuity_new}) is written in a flux-conservative form,
while eq.~(\ref{continuity_old}) is written in a ``flux-balanced'' form,
as it is typical for flux-conservative equations written in
curvilinear coordinates~\cite{Leveque92}. The same is true also for
eqs.~(\ref{momentum_equation_new})--(\ref{energy_equation_new}) and
for the corresponding equations presented in
ref.~\cite{Shibata:2003iy}, which are incorrectly classified as
non flux-conservative.

\subsection{Equation of state}

In whatever coordinate system they are written, the system of
hydrodynamics equations can be closed only after specifying an
additional equation, the equation of state (EOS), which relates the
pressure to the rest-mass density and to the energy density. 
The code has been written to use any EOS, but all the tests
so far have been performed using either an (isentropic) polytropic EOS
\begin{eqnarray}
\label{poly}
p &=& K \rho^{\Gamma}\,, \\
e &=& \rho + \frac{p}{\Gamma-1}\,,
\end{eqnarray}
or an ``ideal-fluid'' EOS
\begin{equation}
\label{id fluid}
p = (\Gamma-1) \rho\, \epsilon \,. 
\end{equation}
Here, $e$ is the energy density in the rest frame of the fluid, $K$
the polytropic constant (not to be confused with the trace of the
extrinsic curvature defined earlier) and $\Gamma$ the adiabatic
exponent. In the case of the polytropic EOS (\ref{poly}),
$\Gamma=1+1/N$, where $N$ is the polytropic index and the evolution
equation for $\tau$ does not need to be solved. In the case of the
ideal-fluid EOS (\ref{id fluid}), on the other hand, non-isentropic
changes can take place in the fluid and the evolution equation for
$\tau$ needs to be solved. Note that the polytropic EOS ~(\ref{poly}) is
isentropic and thus does not allow for the formation of physical
shocks, in which entropy (and internal energy) can be increased
locally (shock heating).

\vspace{0.5cm}
\begin{table}
\begin{center}
  \caption{Equilibrium properties of the initial stellar models. The different
  columns refer respectively to: the ratio of the polar to equatorial coordinate radii
  $r_p/r_e$, the central rest-mass density
  $\rho_c$, the gravitational mass $M$, the rest mass $M_0$, the
  circumferential equatorial radius $R_e$, the angular velocity
  $\Omega$, the maximum angular velocity for a star of the same
  rest mass $\Omega_{_{\rm K}}$, the ratio $J/M^2$ where $J$ is the
  angular momentum, the ratio of rotational kinetic energy to
  gravitational binding energy $T/|W|$. All models have been computed
  with a polytropic EOS with $K=100$ and $\Gamma=2$.\\} 

\begin{tabular}{c|ccccccccc}
\hline
  & $r_p/r_e$ & $\rho_c$ & $M$& $M_0$ & $R_e$  & $\Omega$ & $\Omega_K$ & $J/M^{2}$ &$T/|W|$  \\ 
   	        &          & ($\times 10^{-3}$) &($M_{\odot}$)& ($M_{\odot}$)& & &($\times 10^{-2})$ & & \\ 
\hline
A 	& 1.00  & 1.28  & 1.400 & 1.506 & 9.586  & 0.000 & 3.987 & 0.000 & 0.000 \\
B	& 0.67	& 1.28  & 1.651 & 1.786 & 12.042 & 0.253 & 3.108 & 0.594 & 0.081 
\label{tableone}
	\end{tabular}
\end{center}
\end{table}

\section{Numerical tests}
\label{numtests}

In order to test the stability properties of the new formulation
and compare its accuracy with that of the formulation first
presented in~\cite{Shibata:2003iy} and then used, among others, in~\cite{Montero:2004,Montero2008,Shibata03d,ShibataSekiguchiTakahashi2006}, we
have implemented both of them in Whisky2D. After this paper was
published we were made aware via a private communication~\cite{Liu_p}
 that the numerical approach followed in~\cite{Duez05MHD0} is in practice
very similar to our new formulation, although the version in
cylindrical coordinates of the equations discussed in~\cite{Duez05MHD0} is not in
a flux-conservative form; information about such numerical
implementation was not given in~\cite{Duez05MHD0} and therefore not available to
us at the time this work was written.

The initial data, in particular, has been produced as solution of the
Einstein equations for axisymmetric and stationary
stellar configurations~\cite{Stergioulas95}, using the EOS~(\ref{poly})
with $\Gamma=2$ and polytropic constant $K=100$, in order to produce stellar models
that are, at least qualitatively, representative of what is expected
from observations of neutron stars. Our attention has been restricted
to two illustrative models representing a nonrotating star and a
rapidly rotating star having equatorial and polar (coordinate) radii
in a ratio $r_p/r_e=0.67$. The relevant properties of these stellar
models are reported in Table~\ref{tableone}.

All the numerical results presented hereafter have been obtained with
the following fiducial numerical set-up: the reconstruction of the
values at the boundaries of the computational cells is made using the
PPM method~\cite{Colella84}, while the HLLE algorithm is used as an
approximate Riemann solver~\cite{Harten83}. The lapse function is
evolved with the ``$1+\log$'' slicing condition given by
eq.~(\ref{eq:one_plus_log}), while the shift is evolved using a
version of the hyperbolic $\tilde{\Gamma}$-driver
condition~(\ref{gamma-driver}) in which the advection terms for the
variables $\beta^i,\,B^i$ and $\tilde{\Gamma}^i$ are set to zero. The
time evolution is made with a method-of-line approach~\cite{Toro99}
and a third-order Runge-Kutta integration scheme (our CFL factor is
usually chosen between $0.3$ and $0.5$). A third-order Lagrangian
interpolation is adopted to implement the \textit{``cartoon''}
method. For the matter variables we use ``Dirichlet''boundary
conditions (\textit{i.e.,} the solution at the outer boundary is
always kept to be the initial one), while for the field variables we
adopt outgoing Sommerfeld boundary conditions.

We typically present results at four different resolutions: $\bar{h} =
0.4 M,\, \bar{h}/2,\, \bar{h}/4,\, 3\bar{h}/16$ and $\bar{h}/8$, which
correspond to about $25,\, 50,\, 100,\, 133$ and $200$ points across
the stellar radius, respectively. The computational domain extends to
$20\,M$ both in the $x$ and $z$ directions, and a reflection
symmetry is applied across the equatorial (\textit{i.e.,} $z=0$)
plane. Finally, we remark that in contrast with the interesting
analysis of~\cite{Frauendiener02}, we could not find signs of
numerical instabilities when using the above numerical prescriptions
for either of the two formulations considered.

\subsection{Oscillating Neutron stars: fixed spacetime}

The first set of tests we discuss has been carried out by simulating
relativistic polytropic stars in equilibrium and in a fixed spacetime
(\ie in the Cowling approximation). In this case the Einstein equations are
not evolved and the truncation error is in general smaller because
it is produced uniquely from the evolution of the hydrodynamics equations.

\begin{figure}[h]
  \begin{center}
  \includegraphics[width=0.49\textwidth]{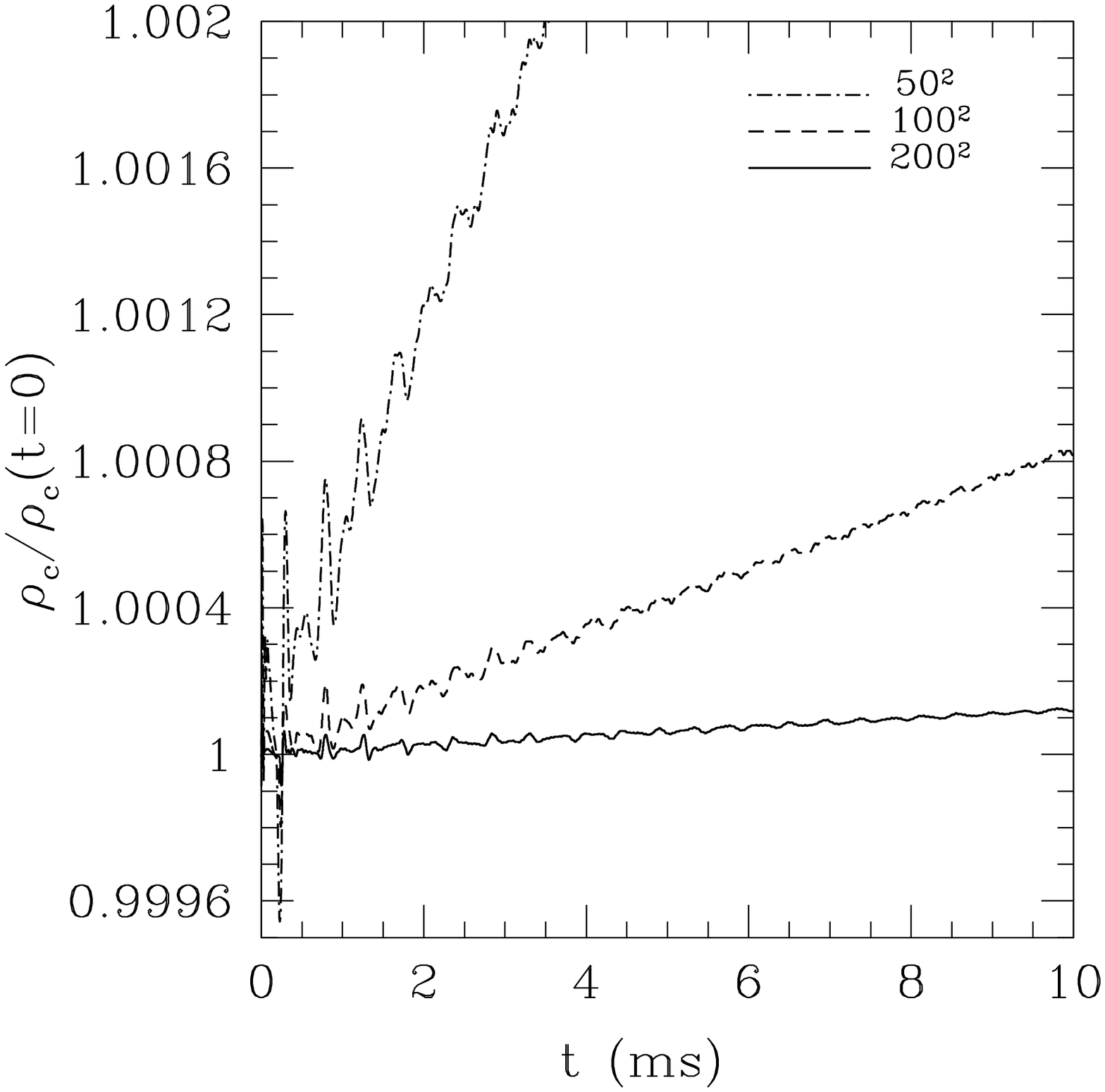}
  \includegraphics[width=0.49\textwidth]{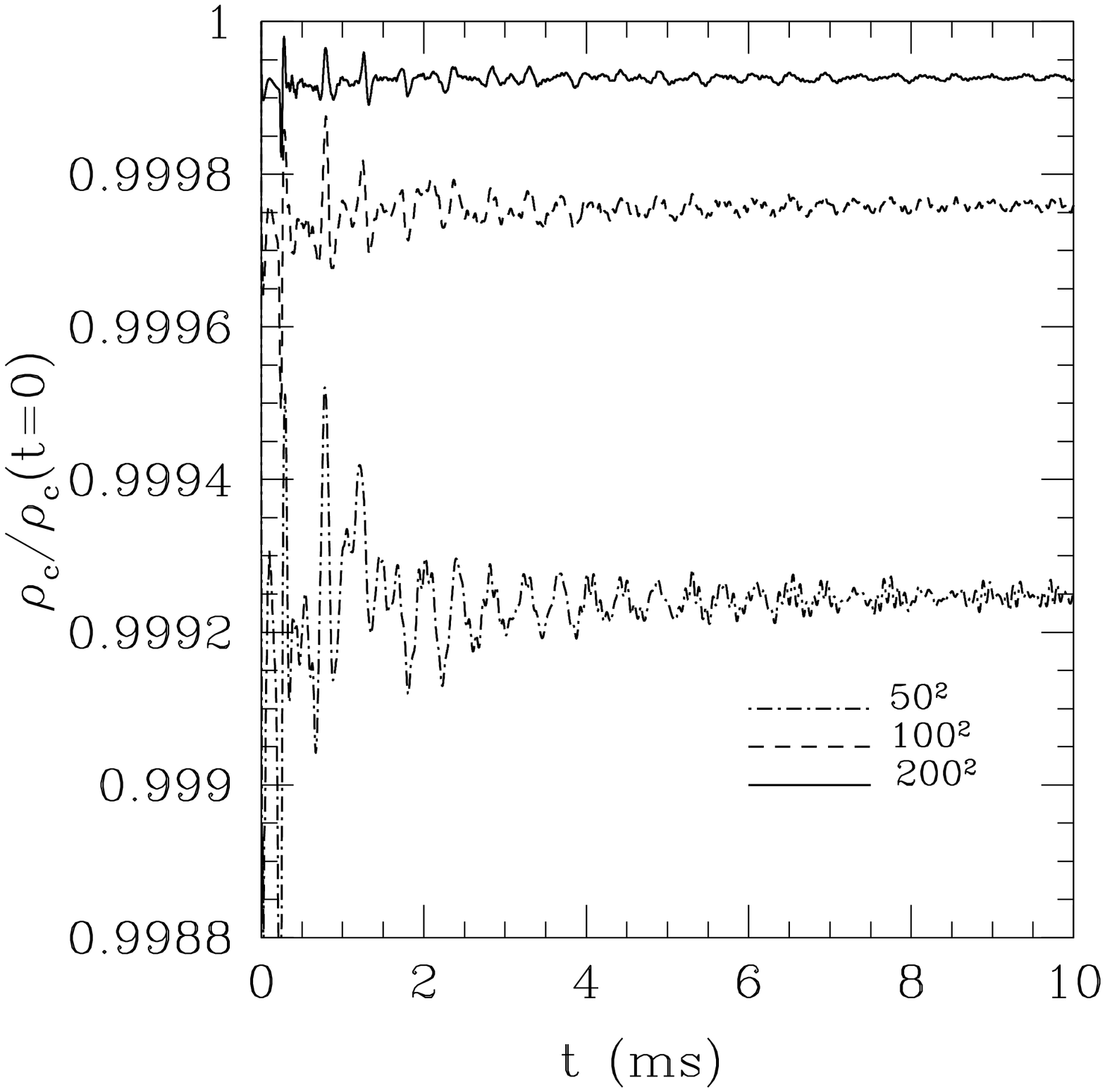}
\end{center}
  \caption{\label{fig:0} Evolution of the central rest-mass density
    for rapidly rotating stars (model B in Table~\ref{tableone})
    evolved within the Cowling approximation. The left panel refers to
    the use of the standard formulation, while the right one to the
    new formulation. Note the different scales in the two panels and
    note that in both cases the amplitude of the oscillations decreases
    with increasing resolution, while keeping the same phase.}
\end{figure}

Although the stars are in equilibrium, oscillations are triggered by
the first-order truncation error at the center and the surface of the
star (our hydrodynamical evolution schemes are only first order at
local extrema). Both the amplitude of the oscillations and the rate of
the secular change in their amplitude converge to zero at nearly
second order with increasing grid
resolution~\cite{Font99,Font02c}. The genuine dynamics produced by the
truncation error can then be studied either when the spacetime is held
fixed (\textit{i.e.,} in the Cowling approximation) or when the
spacetime is evolved through the solution of the Einstein
equations. This is shown in Fig.~\ref{fig:0}, which reports the
evolution of the central rest-mass density for rapidly rotating stars
(model B in Table~\ref{tableone}) evolved within the Cowling
approximation. The left panel refers to the standard formulation,
while the right one to the new formulation. Note that in both cases
the amplitude of the oscillations decreases at roughly second order
with increasing resolution, while keeping the same phase. This is a
clear signature that the oscillations corresponds to proper eigenmodes
of the simulated star. However, the difference of the secular
evolution between the standard formulation and the new one is rather
remarkable.  The latter, in fact, is much more accurate and the
well-known secular increase in the central density is essentially
absent in the new formulation.

Quantities that are particularly useful to assess the accuracy of the
two formulations are the rest mass and the angular momentum which
we compute as~\cite{Yo02a}
\begin{eqnarray}
\fl
&& \hskip -2.0cm
M_{0} = 2\pi\,\int_{V_*} \sqrt{\gamma}\rho W x\,dx\,dz\,, \\ \nonumber \\
\label{Rest_mass}
\fl 
&& \hskip -2.0cm 
J_{z} \,=\,2\pi\, \epsilon_{zj}^k \int_{V} \left(\frac{1}{8\pi}
	\tilde{A}^j_k + x^j S_k + \frac{1}{12\pi} x^j K_{,k} - \frac{1}{16\pi}
        x^j \tilde{\gamma}^{lm}_{,k} \tilde{A}_{lm} \right) e^{6\Phi} x\,dx\,dz\,,
\label{Angmomconservation}
\end{eqnarray}
where $V_*$ is the coordinate volume occupied by the star and $V$ is
coordinate volume of the computational domain.

\begin{figure}[h]
  \begin{center}
  \includegraphics[width=0.49\textwidth]{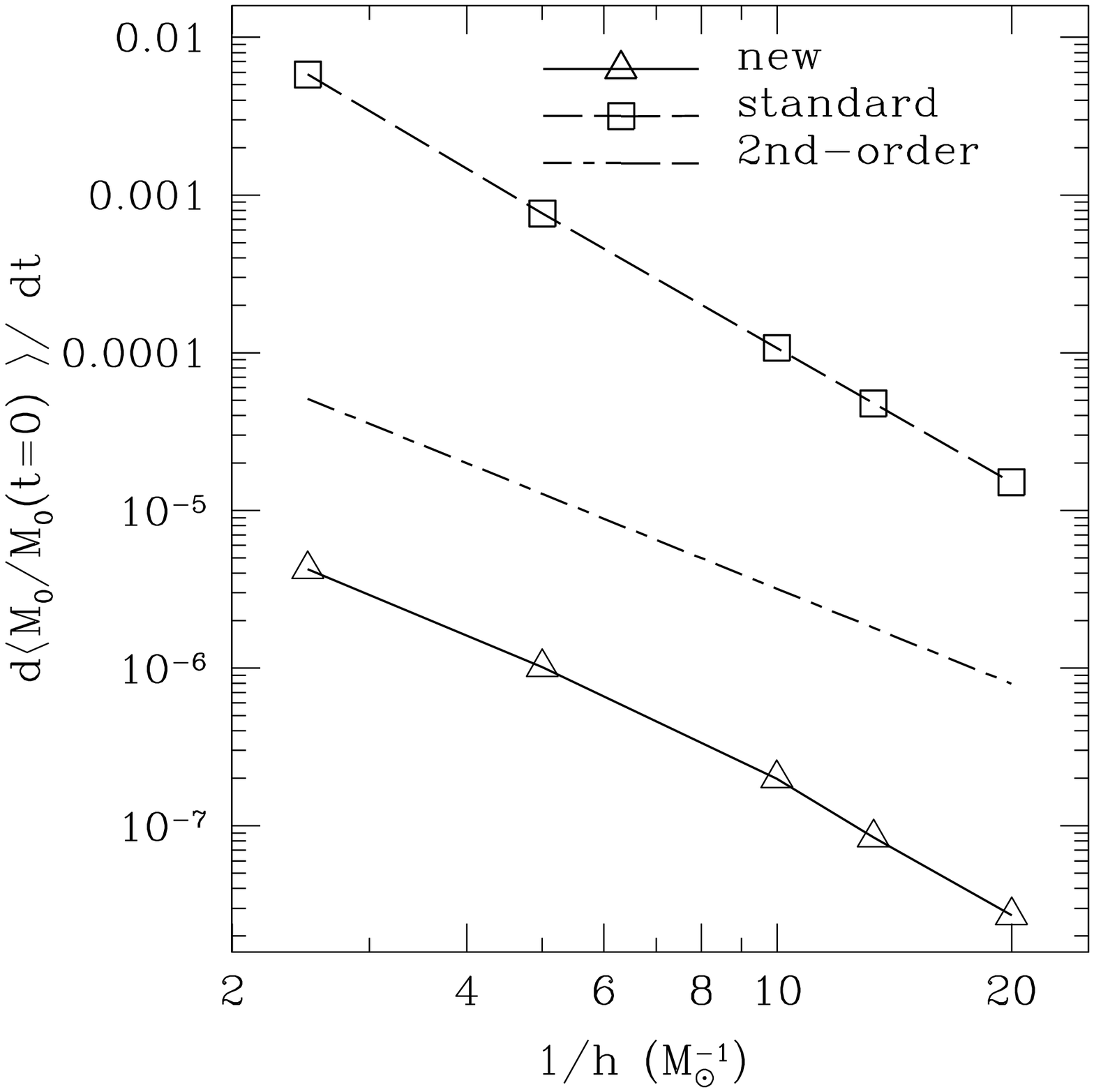}
  \includegraphics[width=0.49\textwidth]{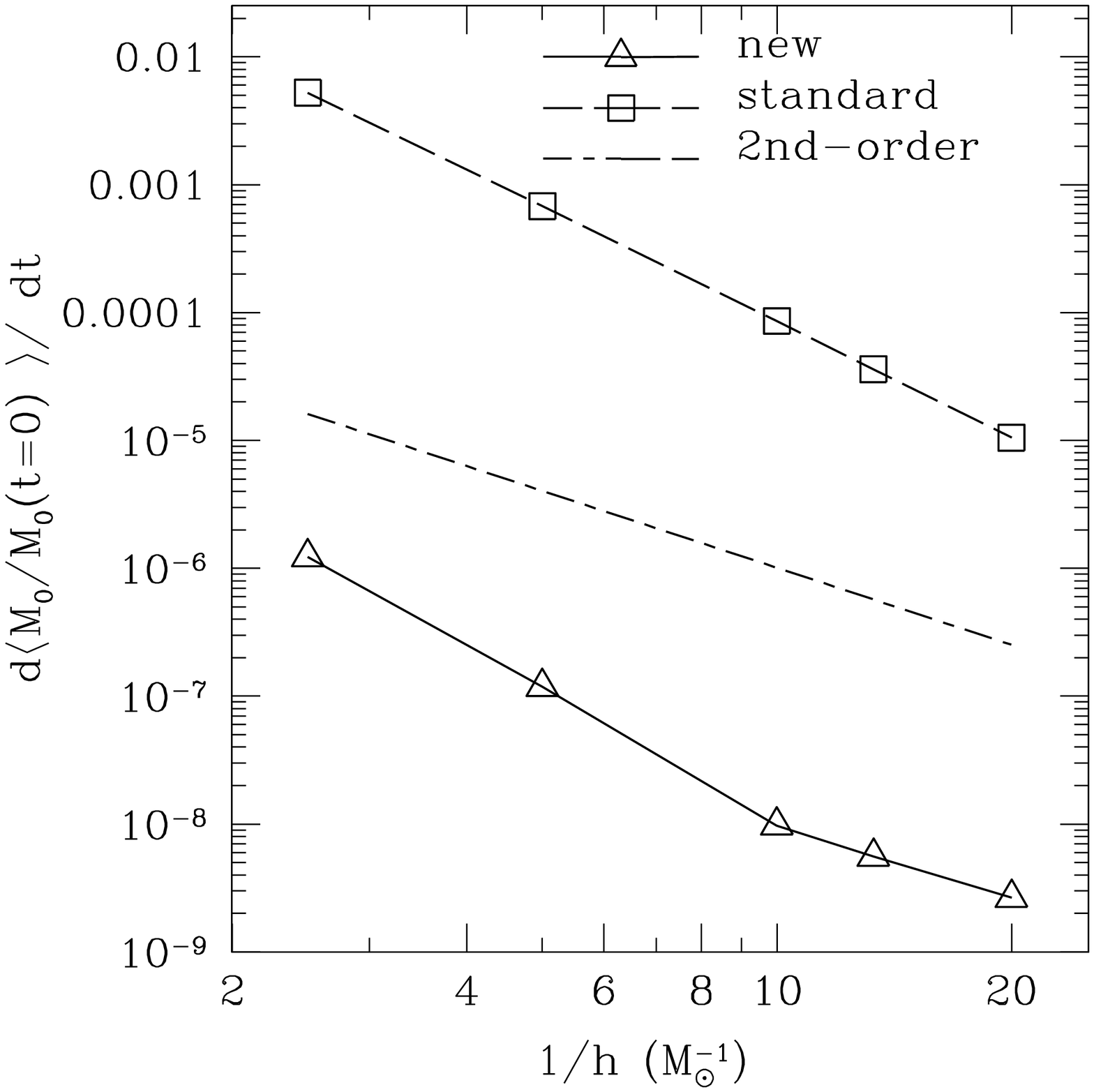}
\end{center}
  \caption{\label{fig:1} Time derivative of the average of the rest mass
    $M_0$, normalized to the initial value $M_0(t=0)$, for evolutions in
    a fixed spacetime (Cowling approximation). The average 
    ${\rm d}\langle M_0/M_0(t=0) \rangle / {\rm d}t$ is computed
    between the initial value and a time $t=25\,{\rm ms}$,
    corresponding to about $30$ oscillations. The left panel refers to a
    nonrotating star (model A in Table~\ref{tableone}), while the
    right panel to a rapidly rotating star (model B in
    Table~\ref{tableone}). Indicated with squares are the numerical
    values obtained with the standard formulation of the hydrodynamics
    equations, while triangles are used for the new one. Also
    indicated with a dot-dashed line is the slope for a
    second-order convergence rate.}
\end{figure}
\begin{figure}[h]
  \begin{center}
  \includegraphics[width=0.49\textwidth]{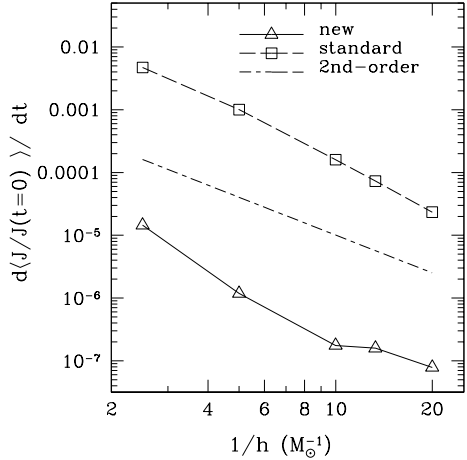}
  \end{center}
  \caption{\label{fig:2} Time derivative of average of the angular
    momentum normalized to the initial value 
    ${\rm d}\langle J/J(t=0) \rangle / {\rm d}t$
    (\textit{cf.}, Fig.~\ref{fig:1}) for a
    rapidly rotating star (model B of Table~\ref{tableone}). Indicated
    with squares are the numerical values obtained with the standard
    formulation of the hydrodynamics equations, while triangles are
    used for the new one; a dot-dashed line is the slope for a
    second-order convergence rate.}
\end{figure}

Figure~\ref{fig:1} shows the dependence on the inverse of the
resolution of the error in the conservation of the rest mass for a
nonrotating model as computed in the Cowling approximation (left
panel) or in a fully dynamical simulation (right panel). Since the
evolution of the rest mass shows, in addition to a secular evolution,
small oscillations (\textit{i.e.,} of $\sim 3\times 10^{-9}$ for the
highest resolution and of $\sim 3\times 10^{-6}$ for the lowest
resolution) the calculation of the rest mass at a given time can be
somewhat ambiguous. To tackle this problem and to avoid the
measurement to be spoiled by the oscillations, we perform a
linear fit of the evolution of $M_0$, normalized to the initial value
$M_0(t=0)$, between the initial value and a time $t=25\,{\rm ms}$
(corresponding to about $30$ oscillations) and we take as the time derivative
of the mass the coefficient of the linear fit: ${\rm d}\langle M_0/M_0(t=0) \rangle / {\rm d}t$. 
Fig.~\ref{fig:1}, in particular, reports in a logarithmic
scale ${\rm d}\langle M_0/M_0(t=0) \rangle / {\rm d}t$ as a function of
the inverse of the resolution $h$. Indicated with squares are the
numerical values obtained with the standard formulation of the
hydrodynamics equations, while triangles are used for the new
one. Also indicated with a long-short-dashed line is the slope for a
second-order convergence rate.

Note that although we use a third-order method for the reconstruction
(namely, PPM), we do not expect third-order convergence. This is also
due to the fact that the reconstruction schemes are only first-order
accurate at local extrema ({\it i.e.}  at the centre and at the
surface of the star), thus increasing the overall truncation
error. Similar estimates were obtained also using the \texttt{Whisky}
code in 3D Cartesian coordinates~\cite{Baiotti03a,Baiotti04}.

Clearly both the new and the standard methods provide a convergence
rate which is close to two. However, and this is the most important
result of this work, the new method yields a truncation error which is
several orders of magnitude smaller than the old one. More
specifically, in the case of the rest mass, the conservation is more
accurate of about four orders of magnitude. We believe that this is
essentially due to the rewriting of the source terms in the
flux-conservative formulation which in the new formulation does not
have any coordinate-singular term ({\it i.e.}  $\propto 1/x$).

Note also that, because the new formulation is intrinsically more
accurate, it also suffers more easily from the contamination of errors
which are not directly related to the finite-difference operators.
[The one made in the calculation of the integral ~(\ref{Rest_mass}) is
a relevant example but it is not the only one]. This may be the reason why, in
general, at lower resolutions the new formulation has convergence rate
which is not exactly two and appears over-convergent (see right panel
of Fig.~\ref{fig:1}). However, as the resolution is increased and the
finite-difference errors become the dominant ones, a clearer trend in
the convergence rate is recovered.

Another way of measuring the accuracy of the two formulations is via
the comparison of the evolution of the angular momentum. While this
quantity is conserved to machine precision in the case of a
nonrotating star, this does not happen for rotating stars and the
error can be of a few percent in the case of very low resolution and
of a very rapidly rotating star. This is shown in Fig.~\ref{fig:2}
for the stellar model B of Table~\ref{tableone} and it reports in a
logarithmic scale the time derivative of the average of the angular momentum
$J$ normalized to the initial value $J(t=0)$. In analogy with
Fig.~\ref{fig:1}, in order to remove the small-scale oscillations we
first perform a linear fit of the evolution of $J$ between the initial
value and a time $t=25\,{\rm ms}$ and take the coefficient of the fit as 
the time derivative of the angular momentum: 
${\rm d}\langle J/J(t=0) \rangle / {\rm d}t$. Indicated with squares are
the numerical values obtained with the standard formulation of the
hydrodynamics equations, while triangles are used for the new one; a
dot-dashed line shows the slope for a second-order convergence rate.

It is simple to recognize from Fig.~\ref{fig:2} that also for the
angular momentum conservation the new formulation yields a truncation
error which is two or more orders of magnitude smaller, with a clear
second-order convergence being recovered at sufficiently high
resolution.

\begin{figure}[h]
  \begin{center}
  \includegraphics[width=0.49\textwidth]{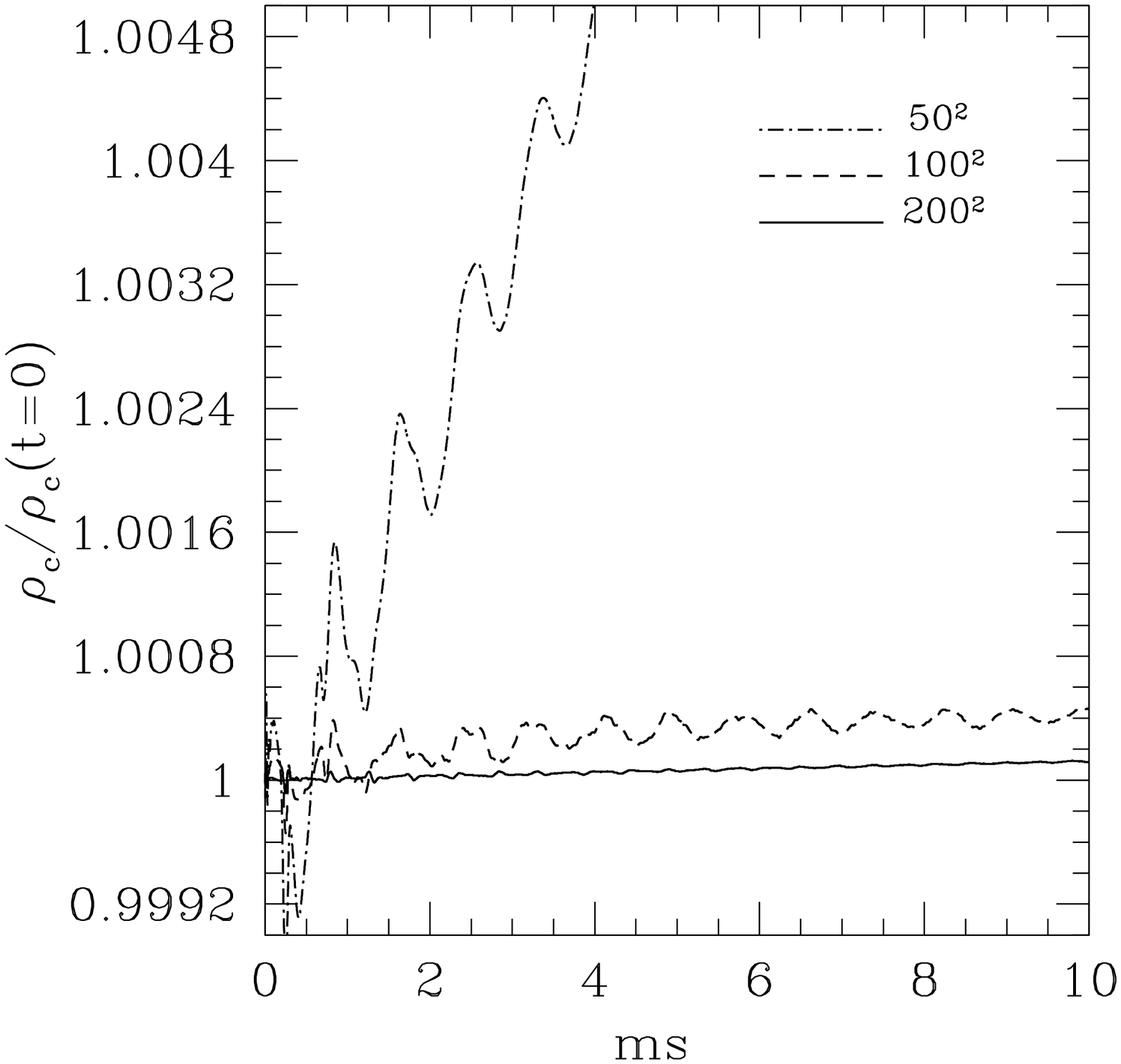}
  \includegraphics[width=0.49\textwidth]{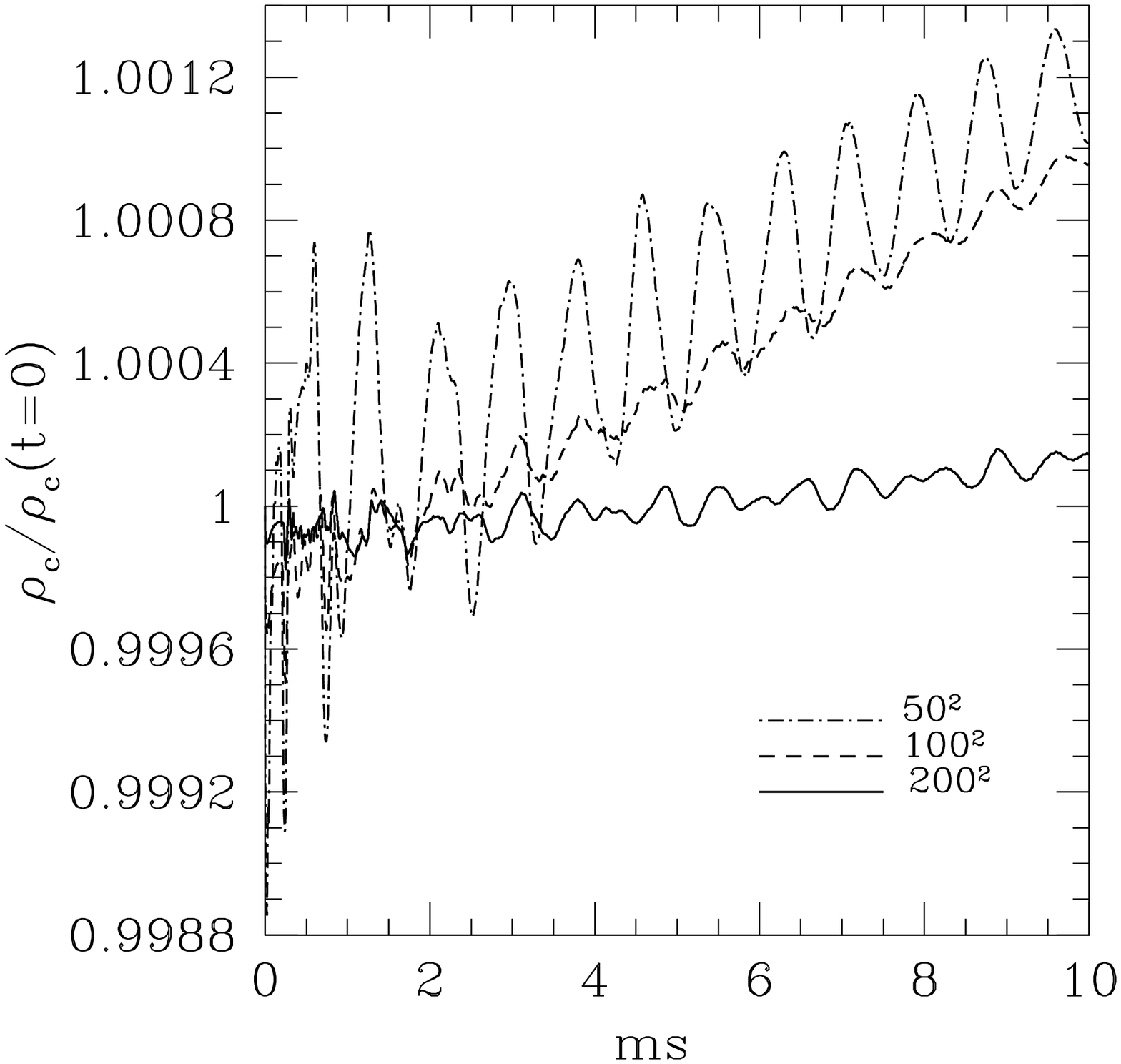}
\end{center}
  \caption{\label{fig:2.5} The same as in Fig.~\ref{fig:0} but for a
    full-spacetime evolution. The left panel refers to the
    standard formulation, while the right one to the new
    formulation. Note the different scale between the two
    panels.}
\end{figure}
\subsection{Oscillating Neutron stars: dynamical spacetime}

\begin{figure}[h]
  \begin{center}
    \includegraphics[width=0.49\textwidth]{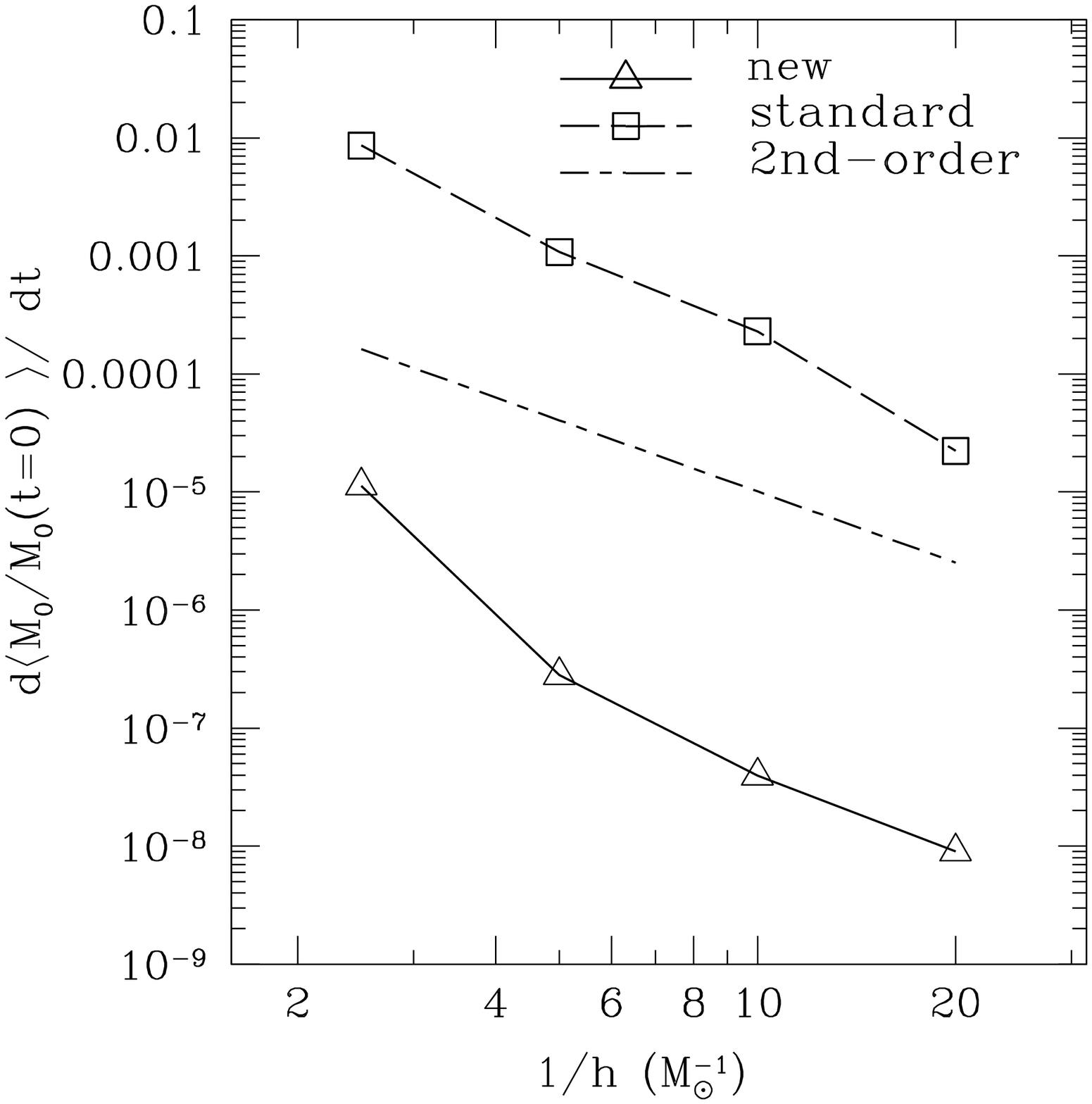}
    \includegraphics[width=0.49\textwidth]{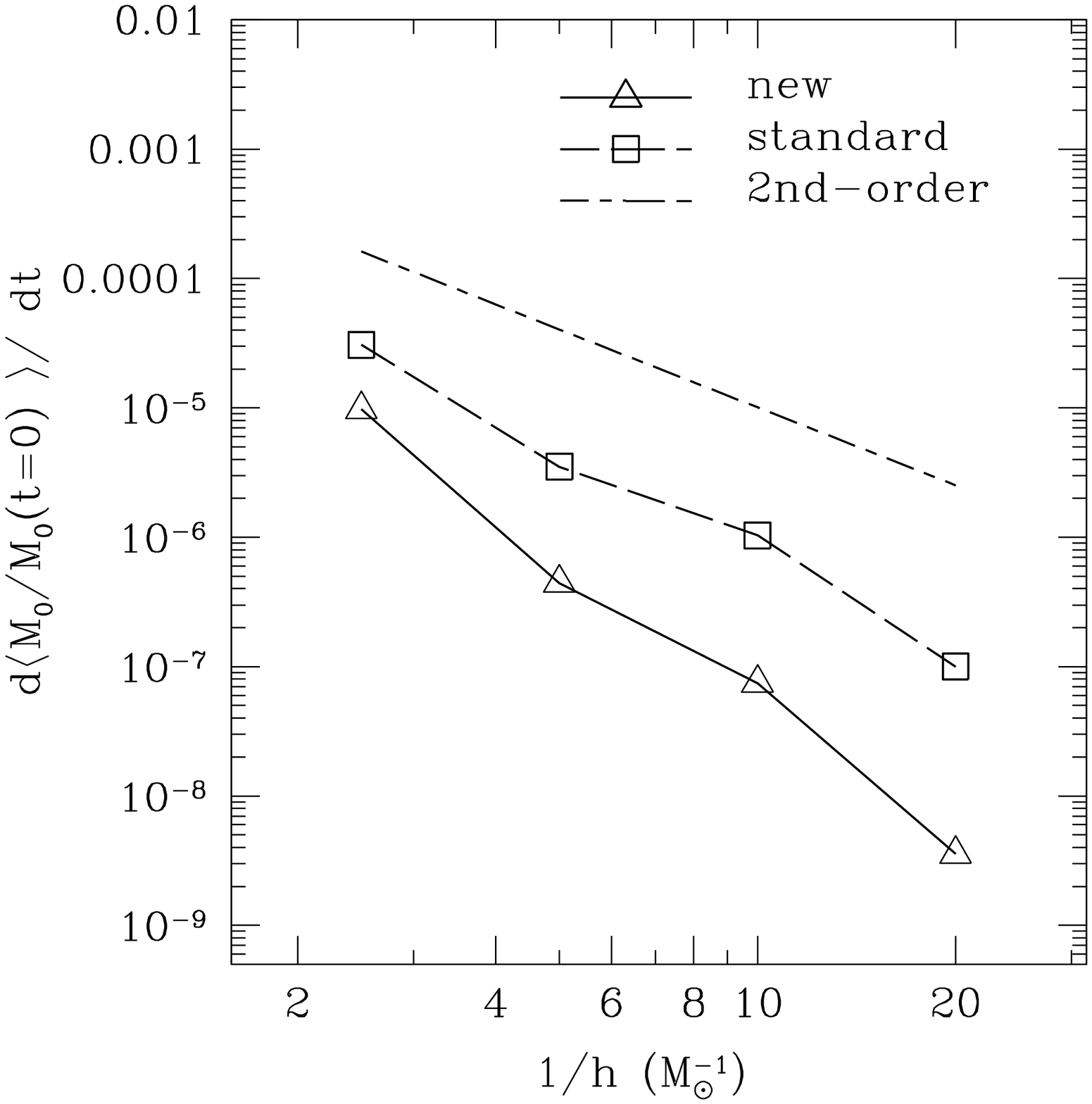}
  \end{center}
  \caption{\label{fig:3} The same as in Fig.~\ref{fig:1}, but for
    full-spacetime evolutions. The left panel refers to a nonrotating
    star (model A in Table~\ref{tableone}), while the right panel to a
    rapidly rotating star (model B in Table~\ref{tableone}).}
\end{figure}

\begin{figure}[h]
  \begin{center}
    \includegraphics[width=0.49\textwidth]{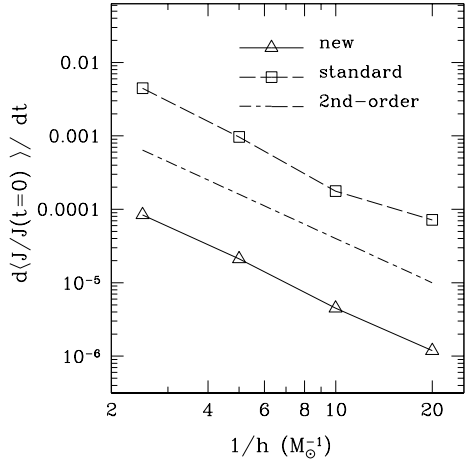}
  \end{center}
  \caption{\label{fig:4} The same as Fig.~\ref{fig:2} but for a
    rapidly rotating star evolved in a dynamical spacetime.}
\end{figure}

Also the second set of tests we discuss is based on the evolution of
relativistic polytropic stars in equilibrium, but now the evolution is performed
in a dynamical spacetime, thus with the coupling of Einstein and
hydrodynamics equations. The truncation error in this case is given by
the truncation error coming from the solution of both the field
equations and the hydrodynamics equations. The results of our
calculations are summarized in Figs.~\ref{fig:2.5}--\ref{fig:4},
which represent the equivalents of Figs.~\ref{fig:0}--\ref{fig:2} for
full-spacetime evolutions. Because the results are self-explanatory
and qualitatively similar to the ones discussed for the evolutions
with fixed spacetimes, we will comment on them only briefly.


In particular, Figs.~\ref{fig:3}--\ref{fig:4} highlight that while the
overall truncation error in dynamical spacetimes is essentially
unchanged for the standard formulation, it has increased in the case
of the new formulation. This is particularly evident at very low
resolutions, where the new formulation seems to be
hyper-convergent. However, despite a truncation error which is larger
than the one for fixed spacetimes, the figures also indicate that the
new formulation does represent a considerable improvement over the
standard one and that its truncation error is at least two orders of
magnitude smaller. Most importantly, the conservation properties of
the numerical scheme have greatly improved and the secular increase in
the rest mass, is also considerably suppressed. This is clearly shown
in Fig.~\ref{fig:2.5}, where the secular increase is suppressed
almost quadratically with resolution. More precisely, for both approaches 
the growth rate of the central rest-mass density for the coarse resolution is $\sim 12$ times
larger than the corresponding one for the high resolution. However, at
the highest resolution, 
the growth rate for the standard formulation is $\sim 10$ times larger than the one of the 
new formulation.

\subsection{Calculation of the eigenfrequencies}

As mentioned in the previous Section, although in equilibrium, the
simulated stars undergo oscillations which are triggered by the
nonzero truncation error. It is possible to consider these oscillations not as 
a numerical nuisance, on the contrary it is possible to exploit them to
perform a check on the consistency of a full nonlinear evolution with a small
perturbation (the truncation error) with the predictions of perturbation
theory~\cite{Font99,Font02c}. Furthermore, when used in conjunction with
highly accurate codes, these oscillations can provide important
information on the stellar oscillations within regimes, such as those
of very rapid or differential rotation, which are not yet accessible
via perturbative calculations~\cite{Dimmelmeier06}.

\begin{figure}[h]
  \begin{center}
   \includegraphics[width=0.49\textwidth]{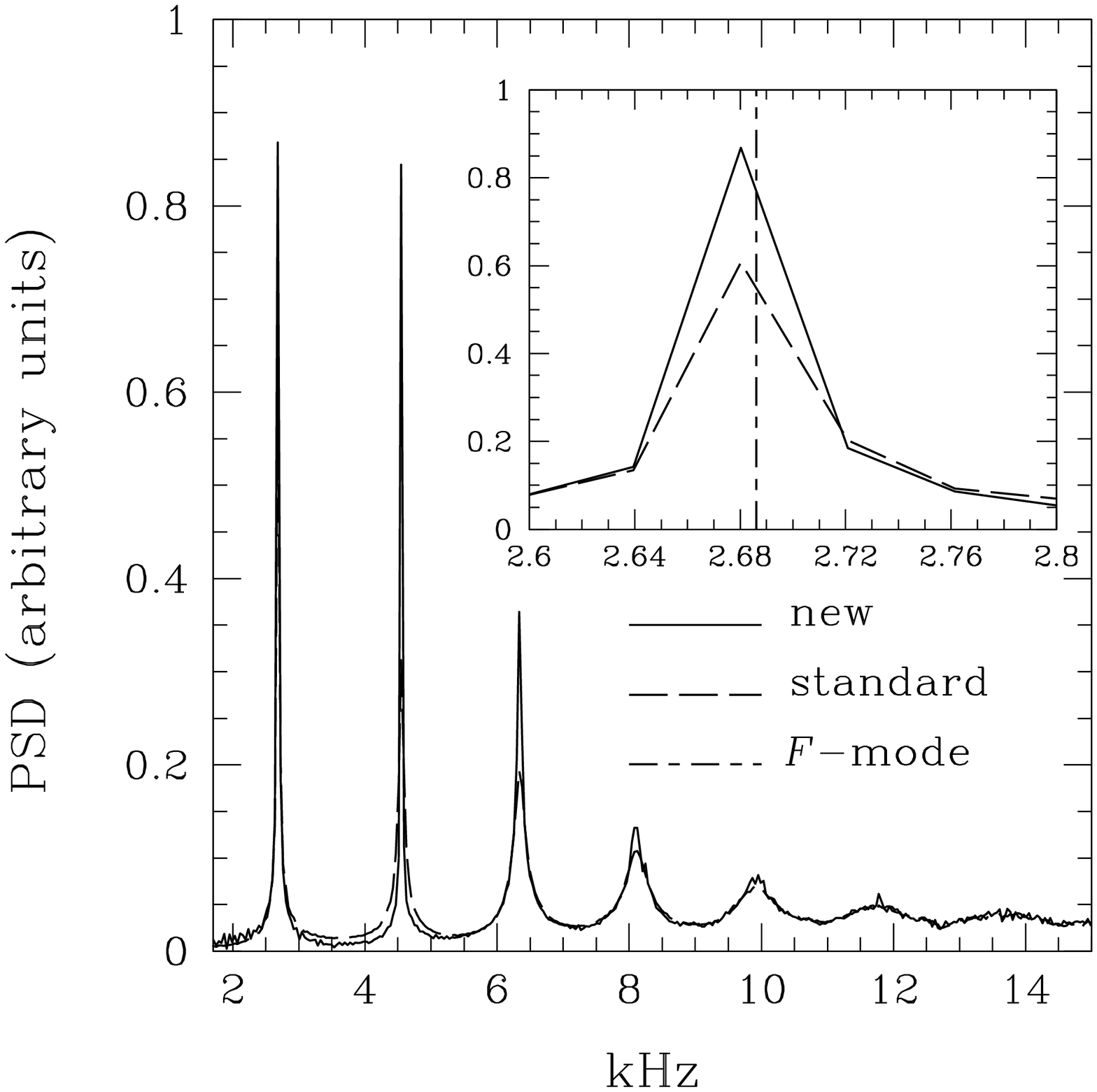}
   \includegraphics[width=0.49\textwidth]{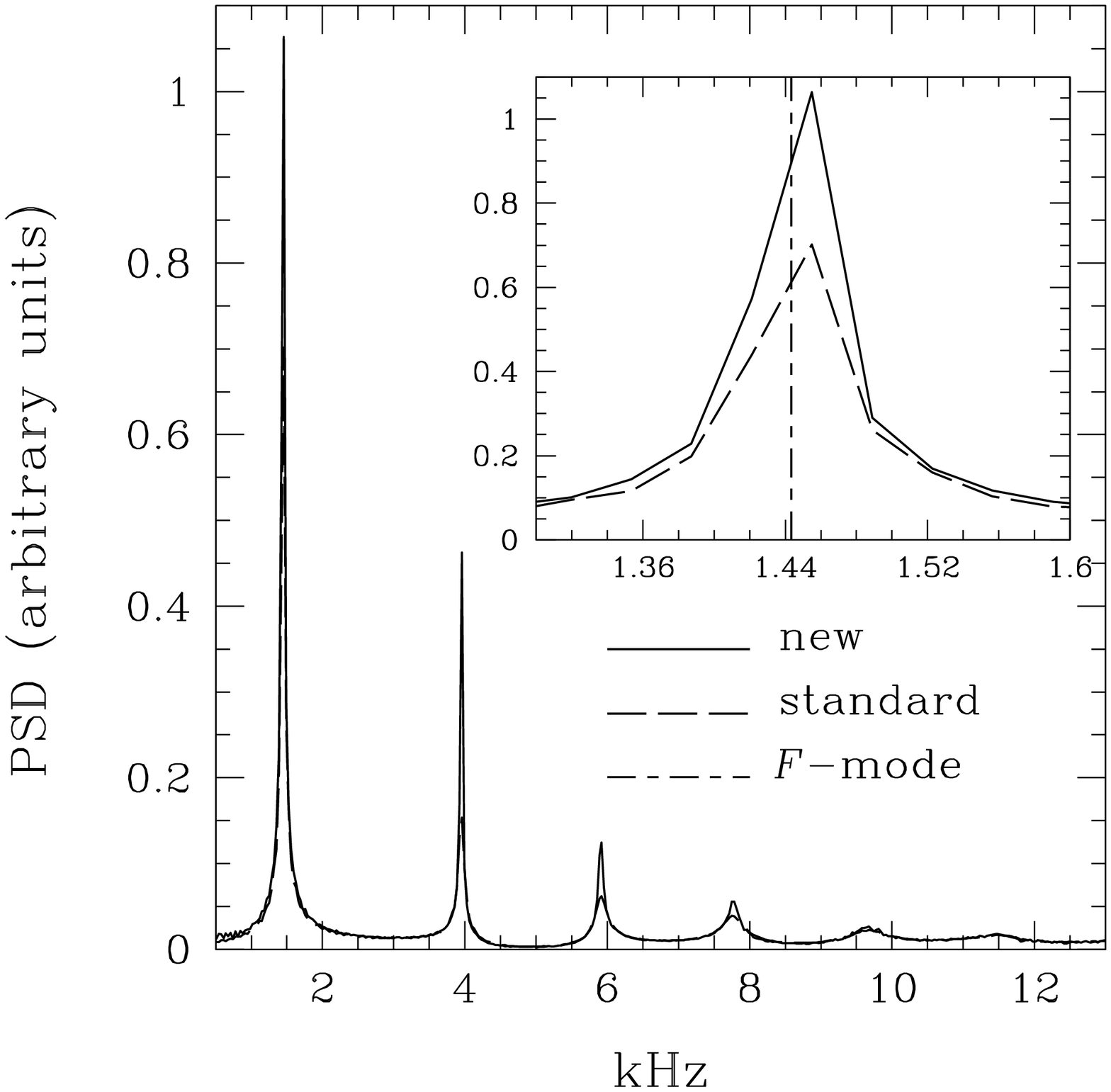}
  \end{center}
  \caption{\label{fig:5} Power spectral density (in arbitrary units)
    of the maximum rest-mass density evolution in the new and standard
    formulation (solid and dashed lines, respectively). The
    simulations are relative to a nonrotating star (model A in
    Table~\ref{tableone}) with the left panel referring to an
    evolution with a fixed spacetime and the right one to an
    evolution with a dynamical spacetime. The spectra are calculated
    from the simulations at the highest resolution and cover $25
    \,{\rm ms}$ of evolution. For both panels the insets show a
    magnification of the spectra near the $F$-mode and the comparison
    with the perturbative estimate as calculated with the numerical
    code described in ref.~\cite{yoshida01}.}
\end{figure}

In this Section we use such oscillations, and in particular the
fundamental $\ell = 0$ quasi-radial $F$-mode, to compare the accuracy
of the two formulations against the perturbative predictions. This is
summarized in Fig.~\ref{fig:5} which reports the power spectral
density (in arbitrary units) of the maximum rest-mass density
evolution (\textit{cf.}, Figs.~\ref{fig:0} and~\ref{fig:2}) in the new
and standard formulation (solid and dashed lines, respectively). The
simulations are relative to a nonrotating star (model A in
Table~\ref{tableone}) with the left panel referring to an evolution
with a fixed spacetime, while the right one to an evolution with a
dynamical spacetime. The specific spectra shown are calculated from
the simulations at the highest resolution and cover an interval of $25
\,{\rm ms}$.  It is quite apparent that the two formulations yield
spectra which are extremely similar, with a prominent $F$-mode at
about $2.7\, {\rm kHz}$ and $1.4\, {\rm kHz}$ for the fixed and
dynamical spacetime evolutions, respectively. The spectra also show
the expected quasi-radial overtones at roughly multiple integers of
the $F$-mode, the first of which has a comparable power in the case of
Cowling evolution, while it is reduced of about $50\%$ in the full
spacetime evolution. Indeed, the spectra in the two formulations are
so similar that it is necessary to concentrate on the features of the
$F$-mode to appreciate the small differences. These are shown in the
insets of the two panels which report, besides a magnification of the
spectra near the $F$-mode, also the perturbative estimate $F_{\rm
  pert}$, as calculated with the perturbative code described in
ref.~\cite{yoshida01}.

\begin{figure}[h]
  \begin{center}
   \includegraphics[width=0.49\textwidth]{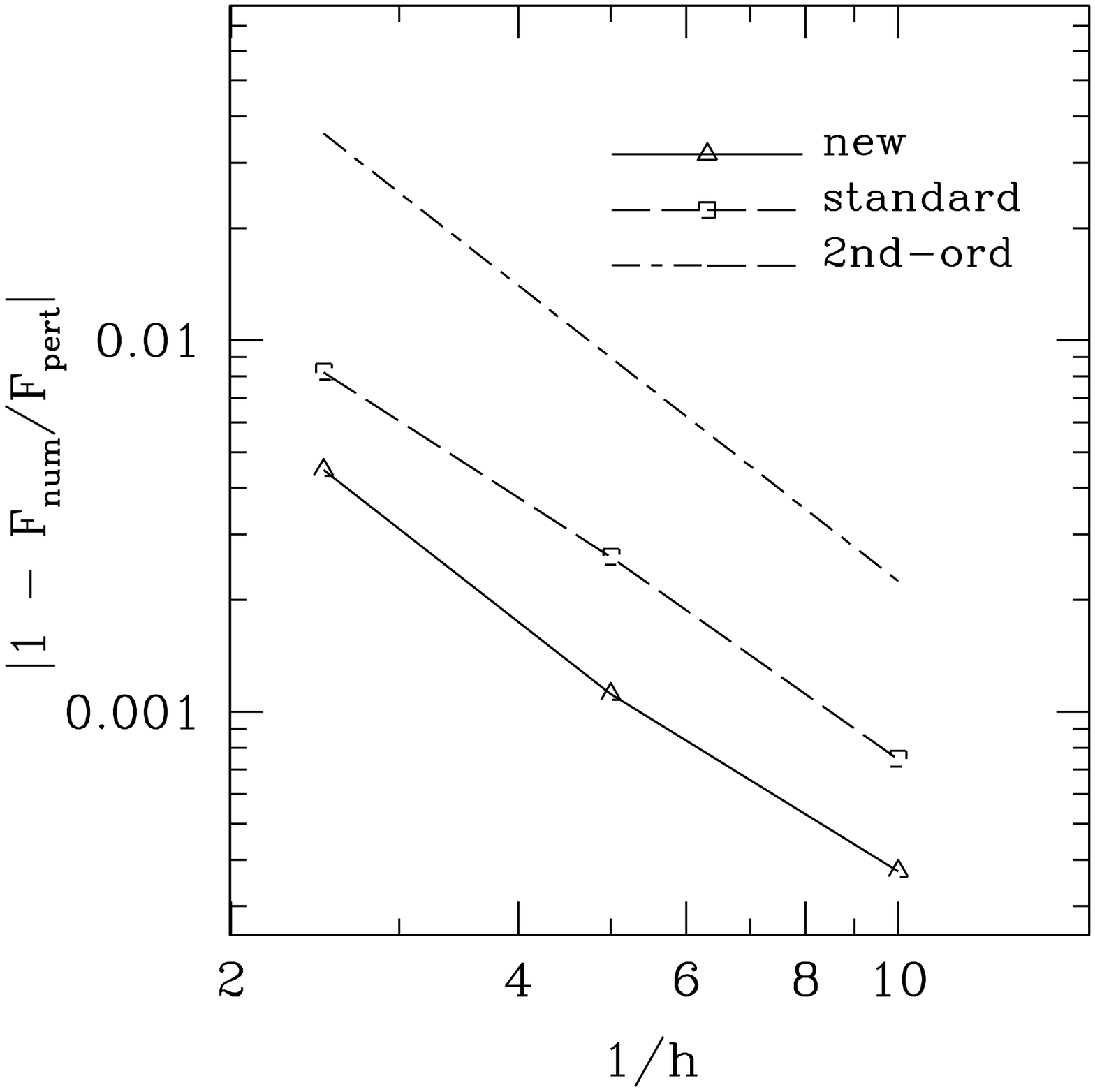}
   \includegraphics[width=0.49\textwidth]{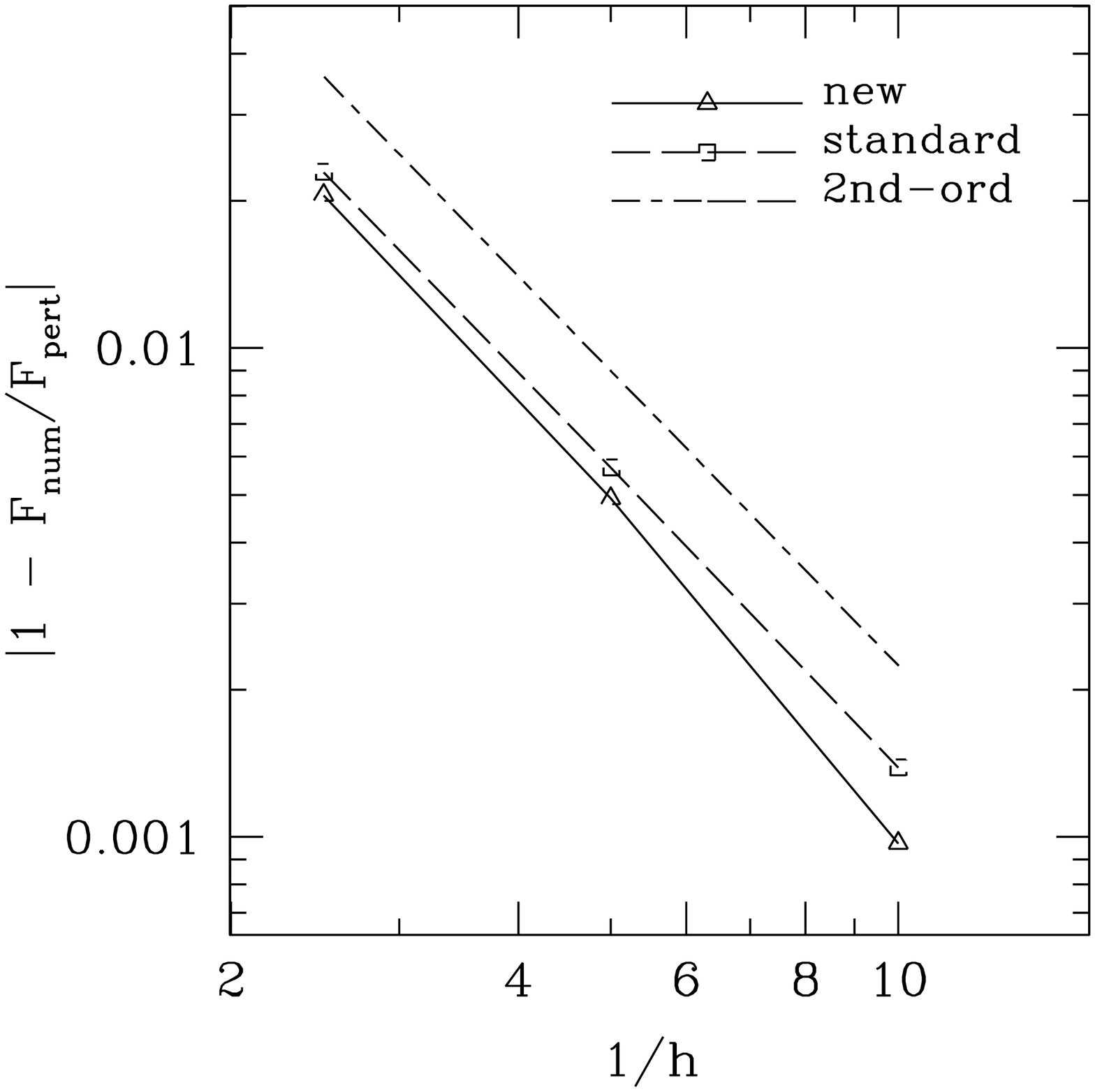}
  \end{center}
  \caption{\label{fig:6} Relative difference between the numerical and
    perturbative eigenfrequencies of the $F$-mode for the two
    formulations (solid lines for the new one and dashed lines for the
    standard one). The differences are computed for different
    resolutions and refer to the nonrotating model A of
    Table~\ref{tableone} when evolved in a fixed spacetime (left
    panel) and in a dynamical one (right panel). Indicated with a
    dot-dashed line is the slope for a second-order convergence rate.}
\end{figure}

To provide a more quantitative assessment of the accuracy with which
the two formulations reproduce the perturbative result we have
computed the eigenfrequency of the $F$-mode, which we indicate as
$F_{\rm num}$, by performing a Lorentzian fit to the power spectrum
with a window of $0.2$ kHz. We remark that it is essential to make use
of a Lorentzian function for the fit as this reflects the expected
functional behaviour and increases the accuracy of the fit
significantly. Shown in Fig.~\ref{fig:6} is the absolute value of the
relative difference between the numerical and perturbative
eigenfrequencies of the $F$-mode, $|1 - F_{\rm num}/F_{\rm pert}|$ for
the two formulations (solid lines for the new one and dashed lines for
the standard one). The differences are computed for different
resolutions with $\bar{h} = 0.4 M,\, \bar{h}/2 $ and $\bar{h}/4$ and refer to the nonrotating mode A of Table~\ref{tableone}
when evolved in a fixed spacetime (left panel) and in a dynamical one
(right panel). Indicated with a dot-dashed line is the slope for a
second-order convergence rate. This helps to see that both
formulations yield an almost second-order convergent measure of the
eigenfrequencies of the $F$-mode, with the new formulation having a
truncation error which is always smaller than the one coming from the
standard formulation. Given the importance of an accurate measurement
of the eigenfrequencies to study the mode properties of compact stars,
we believe that Figs.~\ref{fig:5} and~\ref{fig:6} provide an
additional evidence of the advantages of the new formulation.

Finally, we note that a behaviour similar to the one shown in
Fig.~\ref{fig:5}--~\ref{fig:6} has been found also for rotating stars
although in this case the comparison is possible only for evolutions
within the Cowling approximation since we lack a precise perturbative
estimate of the eigenfrequency for model B of Table~\ref{tableone} for
a dynamical spacetime.

\subsection{Cylindrical Shock Reflection}

One of the most important properties of HRSC schemes is their
capability of handling the formation of discontinuities, such as
shocks, which are often present and play an important role in many
astrophysical scenarios. Tests involving shocks formation are usually
quite demanding and codes that are not flux-conservative can also show
numerical instabilities or difficulties in converging to the exact
solution of the problem. Since both the new and the standard
formulation solve the relativistic hydrodynamics equations as written
in a flux-conservative form, they are both expected to be able to
correctly resolve the formation of shocks, although each with its own
truncation error. In the following test we consider one of such
discontinuous flows and show that the new formulation provides a
higher accuracy with respect to the standard one, stressing once again
the importance of the definition of the conserved variables.

More specifically, we consider a one-dimensional test, first proposed
by~\cite{Marti97}, describing the reflection of a shock wave in
cylindrical coordinates. The initial data consist of a pressureless
gas with uniform density $\rho_0=1.0$, radial velocity
$v^x_0=0.999898$, corresponding to an initial Lorentz factor
$W_0=70.0$ and an internal energy which is taken to be small and
proportional to the initial Lorentz factor, \textit{i.e.}, $\epsilon =
10^{-5} (W_0)$. During the evolution an ideal-fluid EOS (\ref{id
  fluid}) is used with a fixed adiabatic index $\Gamma = 4/3$. The
symmetry condition at $x=0$ produces a compression and generates an
outgoing shock in the radial direction. The analytic solution for the
values of pressure, density, gas and shock velocities are given
in~\cite{Marti97}. From them one can determine the position $x_{_{\rm
    S}}$ of the shock front at any time $\bar{t}$
\begin{equation}
\label{shock_pos}
x_{_{\rm S}} \,=\, \frac{(\Gamma-1)W_0|v^x_0|}{W_0+1} \bar{t} \,.
\end{equation}
This can then be used to compare the accuracy of the two
formulations.

\begin{figure}[h]
   \begin{center}
    \includegraphics[width=0.49\textwidth]{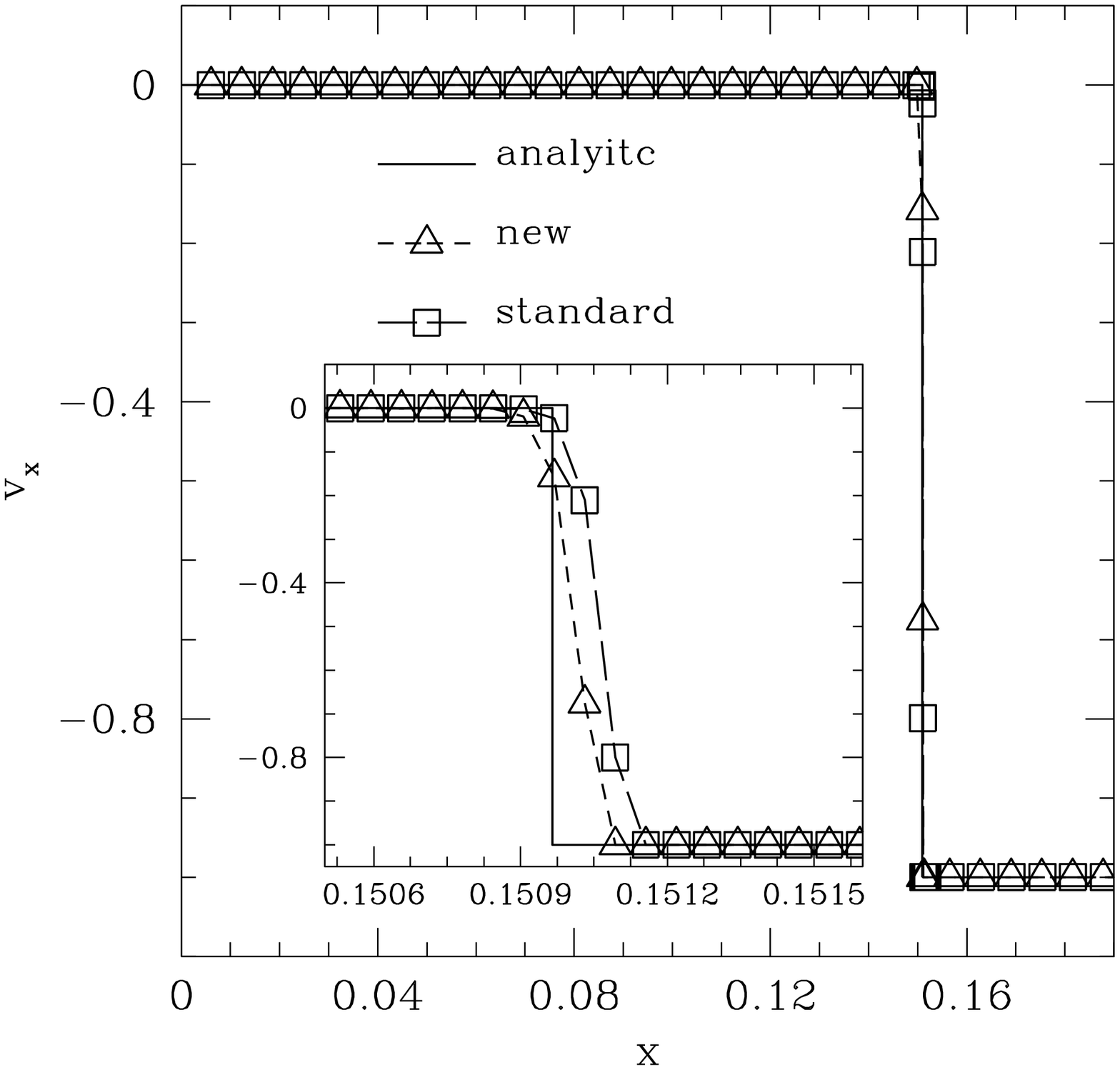}
    \includegraphics[width=0.49\textwidth]{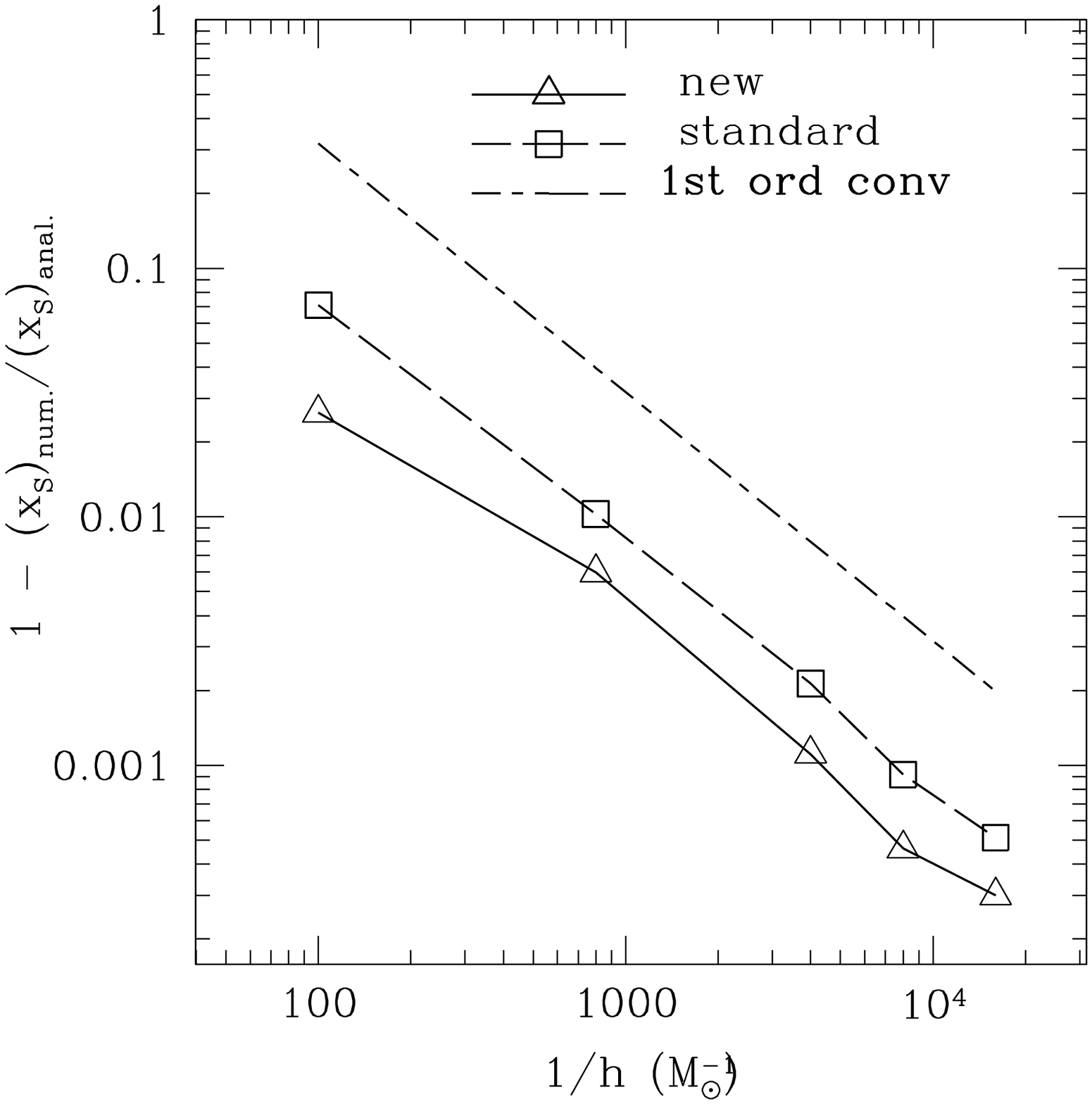}
   \end{center}
   \caption{\label{fig:7} \textit{Left panel:} Comparison of the
     velocity profiles for the two formulations in the solution of the
     axisymmetric shock-tube test with a resolution of $h=6.25\times
     10^{-5}\,M_{\odot}$ The solid line shows the exact position after
     a time $\bar{t} = 0.002262\,{\rm ms}$, while the short-dashed and
     the long-dashed lines represent the solutions with the new and
     the standard formulations, respectively. \textit{Right panel:}
     Comparison of the error in the determination of the position of
     the shock in the two formulations. Note the first-order
     convergence rate as expected for discontinuous flows.}
 \end{figure}

In the left panel of Fig.~\ref{fig:7} we show the value of the radial
component of the velocity $v_x$ as a function of $x$ at a time
$\bar{t}=0.002262\,{\rm ms}$ and for a resolution of
$h/M_{\odot}=6.25\times 10^{-5}$.  The solid line represent the
analytic solution, the short-dashed line the numerical solution
computed with the new formulation and the long-dashed line the one
obtained with the standard formulation. As it is evident from the
inset, the position of the shock is very well captured by both
formulations, but the new one is closer to the exact one at this time.

To compare with the exact prediction given by
expression~(\ref{shock_pos}), we compute the numerical position of the
shock as the middle of the region where the value of the velocity
moves from the pre-shock value $v^+_x$ to the post-shock one $v^-_x$
(in practice, we fit a straight line between the last point of the
constant post-shock state and the first point of the constant
pre-shock state and evaluate the position at which this function has
value $(v^+_x + v^-_x)/2$.). The right panel of Fig.~\ref{fig:7},
shows the relative error $1 - (x_{_{\rm S}})_{\rm num}/(x_{_{\rm
    S}})_{\rm anal}$ in the position of the shock at time $\bar{t}=0.002622\,{\rm
  ms}$ and for five different resolutions: $\bar{h}=0.01\,M_{\odot}$,
$\bar{h}/8$, $\bar{h}/40$, $\bar{h}/80$ and $\bar{h}/160$.  Indicated
with a dashed line is the error computed when using the standard
formulation, while indicated with a solid line is the error coming
from the new formulation. Note that both formulations show a
first-order convergence, as expected for HRSC schemes in the presence
of a discontinuous flow, but, as for the other tests, also in this case
the new formulation has a smaller truncation error. A similar
behaviour is shown also by other quantities in this test but these are
not reported here.

It is useful to note that the difference between the two formulations
in this test is smaller than in the previous ones, being of a factor
of a few only and not of orders of magnitude. We believe this is due
in great part to the fact that, in contrast with what happens for
stars, the solution in the most troublesome part of the numerical
domain (\textit{i.e.} near $x\sim 0$, $z \sim 0$) is not characterized
by particularly large values of the fields or of the fluid variables.
In support of this conjecture is the evidence that at earlier times,
when the shock is closer to the axis, both the absolute errors and the
difference between the two formulations are larger.

\section{Conclusion}

A number of astrophysical scenarios can be very conveniently studied
numerically by assuming they possess and preserve a rotation symmetry
around an axis. Such an assumption reduces the number of spatial
dimensions to be considered and thus the computational costs. This, in
turn, allows for a more sophisticated treatment of the physical and
astrophysical processes taking place and, as a result, for more
realistic simulations.

We have presented a new numerical code developed to solve in Cartesian
coordinates the full set of general relativistic hydrodynamics
equations in a dynamical spacetime and in axisymmetry. More
specifically, the new code solves the Einstein equations by using the
\textit{``cartoon''} method, while HRSC schemes are used to solve the
hydrodynamic equations written in a conservative form. An important
feature of the code is the use of a novel formulation of the equations
of relativistic hydrodynamics in cylindrical coordinates. More
specifically, by exploiting a suitable definition of the conserved
variables, we removed from the source of the
flux-conservative equations those terms that presented coordinate
singularities at the axis and that are usually retained in the
standard formulation of the equations. Despite their simplicity, the
changes made to the standard formulation can produce significant
improvements on the overall accuracy of the simulations with a
truncation error which is often several orders of magnitude smaller.

In order to assess the validity of the new formulation and compare its
accuracy with that of the formulation which is most commonly used in
Cartesian coordinates, we have performed several tests involving the
evolution of oscillating spherical and rotating stars, as well as
shock-tube tests. In all cases considered we have shown that the codes
implementing the two formulations yield the expected convergence rate
but also that the new formulation is always more accurate, often
considerably more accurate, than the standard one.

In view of its simplicity, the new formulation of the equations can be
implemented straightforwardly in codes written using the standard
formulation and we recommend its use for all situations in which an
axisymmetric problem needs to be investigated in full general
relativity and in Cartesian coordinates.

\ack

It is a pleasure to thank Shin'ichirou Yoshida for providing us with
the perturbative eigenfrequencies and Pedro Montero, Olindo Zanotti
and Toni Font for useful discussions. The computations were performed
on the clusters Peyote, Belladonna and Damiana of the AEI. This work
was supported in part by the DFG grant SFB/Transregio~7 and by the
JSPS Postdoctoral Fellowship For Foreign Researchers, Grant-in-Aid for
Scientific Research (19-07803).

\section{Appendix A}

In what follows we recall the essential features of the
\textit{``cartoon''} method for the solution of the field equations in
Cartesian coordinates.  Consider therefore the computational domain to
have extents $0 \le x,z \le d_{max}$ and $-\Delta y \le y \le \Delta
y$, where $d_{max}$ refers to the location of the outer boundary.
Reflection symmetry with respect to the $z=0$-plane can additionally be
assumed. The Einstein equations are then solved only on the
$y=0$-plane with the derivatives in the $y$-direction being computed
with second-order centred stencils using the points at $-\Delta
y,\,0\,,\Delta y$.

Taking into account axisymmetry, the rotation in the $(x,y)$ plane is
defined as
\begin{eqnarray}
R(\phi)_{j}^{i} \,=\, \left(\begin{array}{ccc} \cos({\phi}) &
  -\sin({\phi}) & 0\\ \sin({\phi}) & \cos({\phi}) & 0\\ 0 & 0 & 1\\
	\end{array}\right)\,,
\end{eqnarray}
where $R(\phi)^{-1} \,=\, R(-\phi)$ and the rotation angle is defined
as $\phi = \tan^{-1}(\pm\Delta y / \sqrt{x^2+(\Delta y)^{2}})$.

As commented in the main text, the values of all the quantities on the
$\pm \Delta y$ planes are computed via rotations of the corresponding
values on the $y=0$-planes. More specifically, the components of an
arbitrary vector field $T_{i}$ on the $\pm \Delta y$ planes are
computed via a $\phi$-rotation as
\begin{eqnarray}
T_{x} = T^{(0)}_{x} \cos(\phi) - T^{(0)}_{y} \sin(\phi)\,,\\
T_{y} = T^{(0)}_{x} \sin(\phi) + T^{(0)}_{y} \cos(\phi)\,,\\
T_{z} = T^{(0)}_{z}\,,
\end{eqnarray}
where $T^{(0)}_{i}$ denote the corresponding components at
$(\sqrt{x^{2}+(\Delta{y})^{2}},0,z)$, which are computed via a
Lagrangian interpolation from the neighboring points on the $x$-axis.
Similarly, the components of an arbitrary tensor field $T_{ij}$ tensor
will be computed as
\begin{eqnarray}
T_{xx} = T^{(0)}_{xx} \cos^{2}(\phi) - T^{(0)}_{xy} \sin(2\phi) + T^{(0)}_{yy} \sin^{2}(\phi)\,, \\
T_{xy} = \frac{1}{2}T^{(0)}_{xx} \sin(2\phi) - T^{(0)}_{xy} \cos(2\phi) + \frac{1}{2}T^{(0)}_{yy} \sin(2\phi)\,, \\
T_{yy} = T^{(0)}_{xx} \sin^{2}(\phi) - T^{(0)}_{xy} \sin(2\phi) + T^{(0)}_{yy} \cos^{2}(\phi)\,, \\
T_{xz} = T^{(0)}_{xz} \cos(\phi) - T^{(0)}_{yz} \sin(\phi)\,,\\
T_{yz} = T^{(0)}_{xz} \sin(\phi) + T^{(0)}_{yz} \cos(\phi)\,,\\
T_{zz} = T^{(0)}_{zz}\,.
\end{eqnarray}

\vskip 2.0cm

\section*{References}



\bibliographystyle{iopart-num}
\bibliography{fc_vs_nfc}


\end{document}